%% file: main.tex
\documentclass[letterpaper,twocolumn,10pt]{article}
\usepackage{usenix20}
\input{packages}

\usepackage{tikz}
\usepackage{amsmath}

\newcommand{\circled}[1]{%
    \tikz[baseline=(char.base)]{
        \node[shape=circle,draw,inner sep=0.5pt,fill=black,text=white,minimum size=1em] (char) {#1};
    }%
}

\usepackage{array}
\newcolumntype{L}[1]{>{\raggedright\let\newline\\\arraybackslash\hspace{0pt}}m{#1}}
\newcolumntype{C}[1]{>{\centering\let\newline\\\arraybackslash\hspace{0pt}}m{#1}}
\newcolumntype{R}[1]{>{\raggedleft\let\newline\\\arraybackslash\hspace{0pt}}m{#1}}

\newcommand{\sysname}{\texttt{InternEvo}}
\newcommand{\todo}[1]{\textcolor{red}{TODO: #1}}

\definecolor{lightcyan}{rgb}{0.88, 1.0, 1.0}
\definecolor{lightgray}{rgb}{0.83, 0.83, 0.83}
\begin{document}

\date{}

\newcommand{\papername}{\vspace{-20pt}\sysname:  Efficient Long-Sequence Large Language Model Training via \\ Hybrid Parallelism and Redundant Sharding \vspace{-5pt}}


\title{\Large \bf \papername}

\author{
Qiaoling Chen\textsuperscript{\footnotemark[1] \ 1,2,3},
Diandian Gu\textsuperscript{\footnotemark[1] \ 1,4},
Guoteng Wang\textsuperscript{1},
Xun Chen\textsuperscript{5},
Yingtong Xiong\textsuperscript{1},
Ting Huang\textsuperscript{5}
\\
Qinghao Hu\textsuperscript{1,2,3}, 
Xin Jin\textsuperscript{4},
Yonggang Wen\textsuperscript{2}, 
Tianwei Zhang\textsuperscript{2},
Peng Sun\textsuperscript{1,5}
\vspace{5pt}
\\
{\it \textsuperscript{1}Shanghai AI Laboratory}
\qquad
{\it \textsuperscript{2}Nanyang Technological University}
\qquad
{\it \textsuperscript{3}S-Lab, NTU}
\qquad
\\
{\it \textsuperscript{4}Peking University}
\qquad
{\it \textsuperscript{5}SenseTime Research}
\qquad
}

\maketitle

\input{sections/abstract}
\input{sections/introduction}
\input{sections/background}
\input{sections/searchspace}

\input{sections/design}

\input{sections/experiment}

\input{sections/related}
\input{sections/conclusion}
	\label{lastpage}

\bibliographystyle{unsrt}
\bibliography{reference}
\input{sections/appendix}

\label{allpages}

\end{document}

%% file: packages.tex
\usepackage{xcolor}
\usepackage{xspace}
\usepackage{mathtools} 
\usepackage{amssymb}
\usepackage{bm}
\usepackage{enumitem}
\usepackage{textcomp}

\usepackage{graphicx}
\usepackage{subcaption}
\usepackage{tikz}
\usepackage[labelfont=bf]{caption}
\usepackage{bbding}
\usepackage{amsfonts}

\usepackage{tabularx}
\usepackage{multirow}
\usepackage{booktabs}
\usepackage{makecell}
\usepackage[figuresleft]{rotating}
\usepackage[flushleft]{threeparttable}
\usepackage{colortbl}

\usepackage{algorithmicx}
\usepackage{algorithm}
\usepackage[noend]{algpseudocode}

\algnewcommand\algorithmicinput{\textbf{Input:}}
\algnewcommand\Input{\item[\algorithmicinput]}
\algnewcommand\algorithmicoutput{\textbf{Output:}}
\algnewcommand\Output{\item[\algorithmicoutput]}

\usepackage[utf8]{inputenc} 
\usepackage{inconsolata} 
\usepackage[most]{tcolorbox} 
\usepackage{lipsum}
\usepackage{soul} 
\usepackage{siunitx} 
\usepackage{pifont} 
\usepackage{xurl} 

\usepackage[noindentafter]{titlesec}
\titlespacing\section{0pt}{3pt}{3pt} 
\titlespacing\subsection{0pt}{3pt}{3pt}
\titlespacing\subsubsection{0pt}{5pt}{5pt}

\usepackage[skip=3pt]{caption}


%% file: sections/abstract.tex
\begin{abstract}
Large language models
(LLMs) with long sequences begin to power more and more
fundamentally new applications we use every day. 
Existing methods for long-sequence LLM training is neither efficient nor compatible with commonly-used training algorithms such as FlashAttention.
We design \sysname{} to address these issues. \sysname{} decouples all of the sharding dimensions into a new hierarchical space, and systematically analyzes the memory and communication cost of LLM training. Then, it generates an effective hybrid parallelism strategy. We design a new \textit{selective overlap} mechanism to mitigate the communication overhead introduced by the hybrid parallelism. We also implement memory management techniques to reduce GPU memory fragmentation. Evaluation results show that \sysname{} generates parallelization strategies that match or outperform existing methods in model FLOPs utilization.

\end{abstract}

%% file: sections/introduction.tex
\section{Introduction}

With the emergence of large language models (LLM) in recent years, researchers have investigated and proposed many advanced training methodologies in a distributed way, such as 3D parallelism (including data parallelism\cite{KrizhevskySH12, paszke2019pytorch, li2014scaling, li2014communication}, tensor parallelism~\cite{DeanCMCDLMRSTYN12}, and pipeline parallelism~\cite{GPipe, AthlurSSRK22}), PyTorch FDSP~\cite{PyTorchFSDP}, and automatic parallelization frameworks~\cite{Alpa}. 
Recently, LLMs with long sequences have driven the development of novel applications that are essential in our daily lives, including generative AI~\cite{ni2023recent} and long-context understanding~\cite{beltagy2020longformer, zhou2021document}. With the increased popularity of ChatGPT, long dialogue processing tasks have become more important for chat applications than ever~\cite{touvron2023llama}.
In addition to these scenarios for language processing, Transformer-based giant models also achieve impressive performance in computer vision~\cite{zhang2020span, arnab2021vivit, yuan2021tokens} and AI for science~\cite{bi2023accurate, ai4science}, where inputs with long sequences are critical for complex tasks such as video stream processing~\cite{ruan2022survey} and protein property prediction~\cite{chandra2023transformer}.

\begin{figure}[t]
    \centering
    \includegraphics[width=\linewidth]{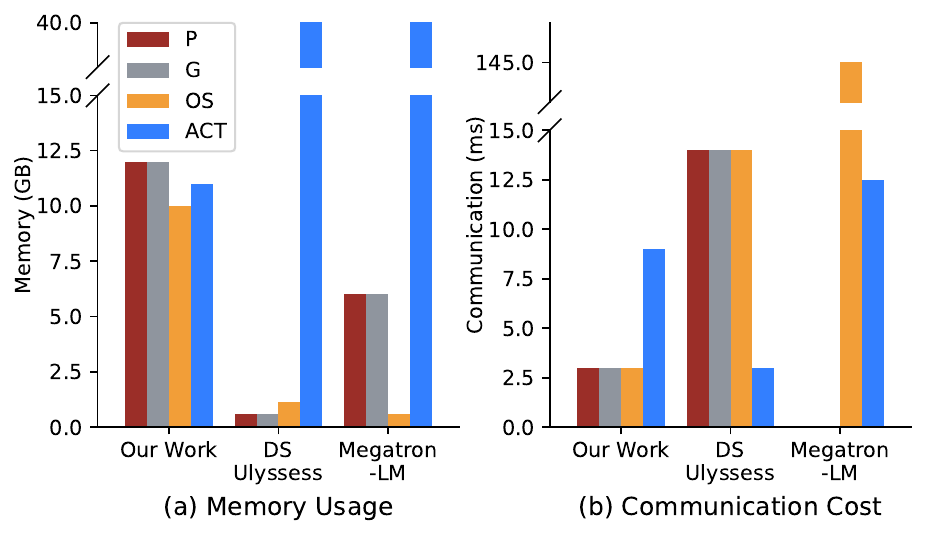}
    \caption{Memory and communication costs of model states (parameters ($P$), gradients ($G$), and optimizer states ($OS$)) and activations (ACT) when training a 7B model with 32k sequence length on 128 GPUs. DS means DeepSpeed here.}
    \label{fig:memandcomm}
\end{figure}

Training LLMs with long sequences requires massive memory resources. 
To fit the memory constraint of GPUs, directly applying 3D parallelism \cite{PipeDream,Megatron-LM} or existing automatic parallelization frameworks \cite{Alpa,FlexFlow,Galvatron,Merak,Unity} is inefficient in training speed~\cite{nvidia3} since these system requires us to hold the whole sequence in one GPU, which limits the length of the input sequence. 
Several methods have been proposed to train LLMs with long sequences while not sacrificing the training speed significantly~\cite{DeepSpeedUlysses, BPT2, li2023lightseq, ColosslaiSP}. However, they suffer from two limitations. 
\emph{First, some studies do not have satisfactory usability}. These parallelization strategies and algorithms~\cite{ColosslaiSP, BPT2, li2023lightseq} compute self-attention in a way that conflicts with FlashAttention~\cite{FlashAttention}. FlashAttention, known for its speed and memory efficiency, is widely utilized in many LLM training~\cite{flashatten1, flashatten2}. 
\emph{Second, the solutions with good usability are still not efficient enough in training performance.} Prior methods such as DeepSpeed Ulysses introduce large inter-node communication overhead~\cite{DeepSpeedUlysses}; other methods like Megatron-LM largely increase the communication demand~\cite{nvidia3}. Additionally, these solutions adopt the \textit{rigid} parallelization strategy. To quantify this, Figure \ref{fig:memandcomm} shows the actual materialized sizes of the memory usage of the model states and activation with the corresponding communication cost for different systems at optimal performance. To alleviate the memory pressure with a sequence length of 32K, both DeepSpeed and Megatron-LM simply partition the model states and activation at the same level (Figure \ref{fig:memandcomm}(a)). Unfortunately, this simplicity incurs the corresponding high communication overhead (Figure \ref{fig:memandcomm}(b)). They do not consider the differences in memory requirements and communication patterns when training with various configurations (e.g. model sizes, batch sizes, sequence lengths), making them not always optimal in different cases. 



To bridge these gaps, 
we propose \sysname{}, a parallelization framework for training Transformer-based LLMs with long sequences. \sysname{} can automatically find an efficient parallelization strategy while different GPU memory management dimensions, including sequence parallelism, traditional 3D parallelism, etc. \sysname{} is also compatible with efficient self-attention optimization algorithms such as FlashAttention. 

However, the characteristics of long-sequence LLM training introduce two challenges for an efficient parallelization strategy. Firstly, there is a trade-off between the computation speed and communication overhead, making it difficult to find an efficient strategy. 
The challenge arises from the fact that reducing the computation load on each GPU by increasing the degree of GPU memory dimensions results in a proportional increase in communication demand, leading to larger communication overhead.
Secondly, the default memory management method for long-sequence LLM training cannot fully utilize GPU memory, which is prone to out-of-memory (OOM) errors. This is caused by two factors. On one hand, parallelization strategies with efficient training speed might require frequent memory allocation and release, introducing memory fragmentation. On the other hand, the architecture of LLaMa-based LLMs requires memory allocation of irregular sizes~\cite{LLaMA}. This also introduces memory fragmentation and limits memory utilization.

To address the first challenge, we redesign the sharding of LLMs and decouple all of the GPU memory management dimensions into a new hierarchical space with four parallel dimensions and three sharding dimensions. In this entirely new search space, we analyze the memory and communication costs of each dimension. Based on the analysis, we design an execution simulator to estimate the final memory and communication cost of every possible parallelization strategy and search for an optimal solution. Figure \ref{fig:memandcomm} shows that \sysname{} performs well in storing more model states in exchange for fewer related communications. We also propose a selective overlap approach to further hide the communication overhead in computation. To address the second challenge, we design a memory pool for unified memory management to mitigate the memory fragmentation problem. This enables high memory utilization while maintaining high training efficiency.

In summary, we make the following contributions:
\begin{itemize}[topsep=0pt]
\setlength{\itemsep}{0pt}
\setlength{\parskip}{0pt}
\setlength{\parsep}{0pt}
\item We summarize the existing memory management dimensions of model training and propose a novel hierarchical space with four parallel dimensions and three sharding dimensions. We systematically analyze the memory and communication costs in this space.
\item We design \sysname{}, an automatic framework for efficient long-sequence LLM training, which automatically finds an efficient parallelization strategy. 
\item We design a fine-grained computation-communication overlap approach to sufficiently hide the communication overhead, and memory management techniques to reduce the GPU memory fragmentation.
\item We implement a system prototype of \sysname{}. Evaluation results on training large models with billions of parameters and up to 256k sequence length show that \sysname{} can bring up to 4.8$\times$ model FLOPs utilization (MFU) improvement compared to competitive baselines.
\end{itemize}

%% file: sections/background.tex
\section{Background}

\begin{table}[htbp]
\centering
\small 
\begin{tabular}{@{}l@{\hspace{0.15cm}}l@{\hspace{0.15cm}}|@{\hspace{0.15cm}}l@{\hspace{0.15cm}}l@{}}
\toprule
$L$   & Number of Transformer Layers & $b$   & Micro-Batch Size \\
$V$   & Vocabulary Size & $n$ & Micro-Batch Number \\
$D$   & Number of Attention Heads & $s_{pp}$ & Pipeline Parallel Size \\
$S$   & Sequence Length (Tokens) & $s_{tp}$ & Tensor Parallel Size \\
$H$   & Hidden Dimension Size & $s_{sp}$ & Sequence Parallel Size \\
$B$   & Global-Batch Size (Tokens) & $s_{dp}$ & Data Parallel Size \\
$M$ & GPU Memory Usage & $s_{ps}$ & Parameter Sharding Size \\
$T^i$  & Time Consumption of Layer-$i$ & $s_{gs}$ & Gradient Sharding Size \\
$N$ & Number of GPUs for the Job & $s_{oss}$ & OS Sharding Size  \\ 
$\Psi$ & Parameter Count of a Layer & $a$ & ACT Recomputation\\ 
\bottomrule
\end{tabular}
\caption{Notations used in this paper.}
\label{tab:transformer-parameters}
\end{table}


We provide a brief introduction to the essential background of Transformer-based LLM architectures and the parallelization techniques for distributed training.

\begin{figure*}[t!]
    \vspace{-4pt}
    \centering
    \includegraphics[width=\linewidth]{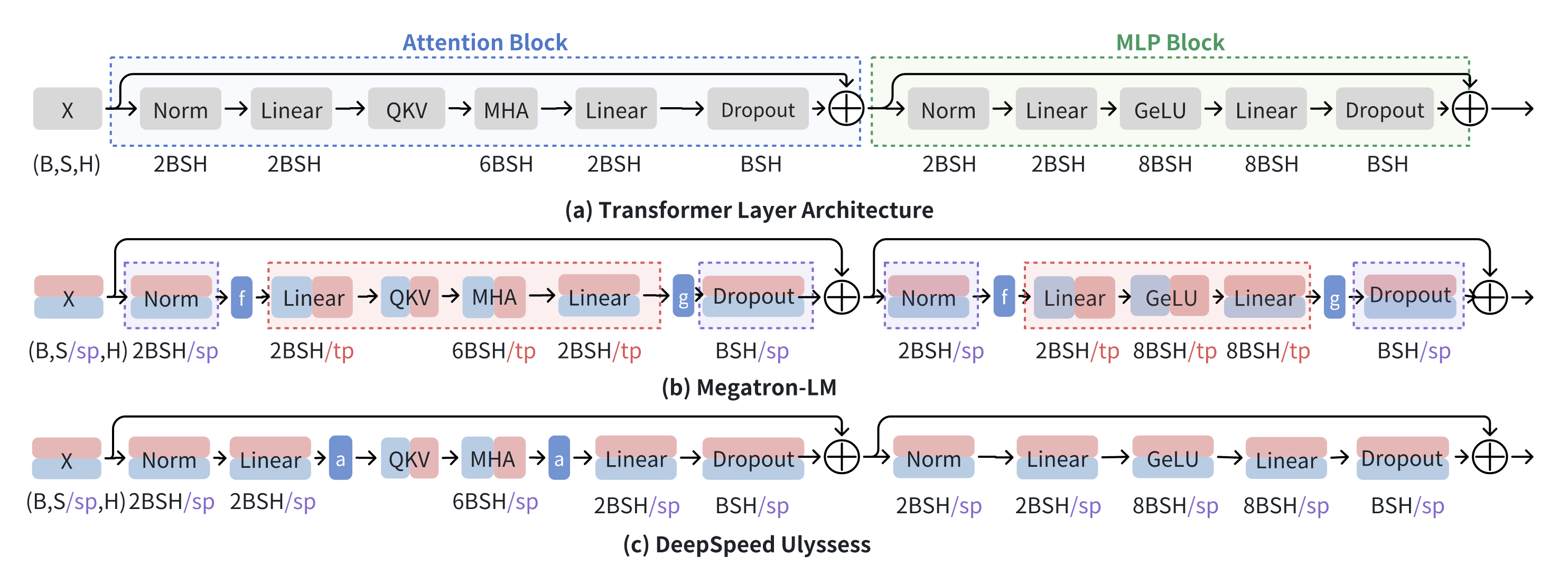}
    \caption{Background of long sequence LLM training. (\textbf{a}) Attention and MLP blocks in the Transformer architecture. The activation size (byte) of each layer is annotated. (\textbf{b}) Hybrid sequence parallelism of Megatron-LM \cite{nvidia3}. $f$ and $g$ represent conjugate communication operators. In the forward pass, $f$: \texttt{all-gather} and $g$: \texttt{reduce-scatter}. In the backward pass, it is the opposite. (\textbf{c}) Sequence parallelism of DeepSpeed  \cite{DeepSpeedUlysses}.  $a$ represents the \texttt{all-to-all} communication operator.}
    \label{fig_background}
\end{figure*}



\subsection{Transformer Architecture}
\label{MemoryFootprint}

LLMs such as GPT-3 \cite{GPT3} and LLaMA \cite{LLaMA} commonly embrace the Transformer architecture \cite{Attention}.  The architectural homogeneity suggests that tailored system optimizations are applicable. As shown in Figure \ref{fig_background} (a), each Transformer layer has an attention block with $D$ attention heads and an MLP layer. Input/output dimensions are $B \times S \times H$ ($B$: micro-batch size, $S$: sequence length, $H$: hidden dimension). We show the memory required to store activations ($ACT$) in a 16-bit floating point format for each element. The fused kernel of FlashAttention \cite{FlashAttention} is used for efficient multi-head attention (MHA) computation.

\subsection{Parallelization in Distributed Training}
\label{3D}

Training LLMs efficiently at scale typically necessitates the combination of various parallelization approaches as follows.

\noindent \textbf{Data Parallelism (DP)} divides the input data into shards and distributes them across GPUs. Each GPU independently conducts computations to obtain gradients, which are subsequently synchronized through all-reduce. 

\noindent \textbf{Tensor Parallelism (TP)} divides parameters along specific dimensions across GPUs to parallelize the training. As shown in Figure \ref{fig_background} (b), Megatron-LM leverages TP to partition linear layers along the row or column dimension, and inserts communication operations to ensure consistent results. 

\noindent \textbf{Pipeline Parallelism (PP)} evenly divides the Transformer layers of a model into multiple stages and distributes them across GPUs. A scheduler splits an input batch into micro-batches and processes forward and backward computations alternately \cite{GPipe,PipeDream}. Two consecutive pipeline stages exchange intermediate data through point-to-point communication. Each pipeline stage exhibits data dependencies on others, resulting in computation stalls, commonly referred to as pipeline \emph{bubbles}.

\subsection{Memory Footprint Conservation}
\label{subsec_memory}

\begin{table}
\small
\centering
\begin{tabular}{@{}l@{\hspace{0.05cm}}|@{\hspace{0.05cm}}c@{\hspace{0.15cm}}c@{\hspace{0.15cm}}c@{\hspace{0.15cm}}c@{\hspace{0.1cm}}||@{\hspace{0.05cm}}c@{\hspace{0.05cm}}|@{\hspace{0.1cm}}l@{}@{}}
\toprule
\textbf{Model} & \textbf{$H$} & \textbf{$L$} & \textbf{$D$} & \textbf{$V$} &\textbf{Data} & \textbf{Memory (Byte)} \\
\midrule
\textbf{7B}  & 4096 & 32 & 32 & 100k & \textbf{$P$}  & $2(\Psi L + 2VH)$  \\
\textbf{13B} & 5120 & 40 & 40 & 100k & \textbf{$G$}  & $2(\Psi L + 2VH)$  \\
\textbf{30B} & 6144 & 60 & 48 & 100k & \textbf{$OS$} & $6(\Psi L + 2VH)$ \\
\textbf{65B} & 8192 & 80 & 64 & 100k & \textbf{$ACT$}  & $2BS(17HL+H+V)$ \\
\bottomrule
\end{tabular}
\caption{Model configurations and memory footprints. $\Psi$ (parameter count of a Transformer layer) = $12H^2+2H$. }
\label{tab:model_used_memory}
\end{table}

Table \ref{tab:model_used_memory} concludes the memory footprint of different model sizes in the context of mixed precision training with the Adam optimizer \cite{Adam}. With the increase of the sequence length, the memory occupied by model states constitutes only a small portion of the total memory. To alleviate the huge memory requirement caused by long sequences, some advanced system techniques are proposed:

\noindent \textbf{Zero Redundancy Optimizer (ZeRO)} \cite{ZeRO}. ZeRO reduces redundant memory usage in DP by sharding model states. Specifically, ZeRO-1 partitions optimizer states ($OS$) across GPUs, allowing each GPU to store only one shard. ZeRO-2 further shards gradients ($G$), and ZeRO-3 (also implemented in FSDP\cite{PyTorchFSDP}) further shards model parameters ($P$). ZeRO-1 and ZeRO-2 will not introduce extra communication. When using ZeRO-3, before conducting forward and backward computations, each GPU participates in an \texttt{all-gather} operation to collect parameters and construct complete model weights.

\noindent \textbf{Activation Recomputation}. It reduces activation memory consumption by checkpointing input activations in some layers, discarding intermediate activations within these layers, and recomputing them through an additional forward pass during the backward pass \cite{chen2016training}. It efficiently reduces the activation memory. However, the recomputation introduces an approximately $30\%$ increase in computation overhead \cite{nvidia3}. 


\subsection{Long-sequence Training}
\label{subsec:sp}
Processing long sequences is crucial for supporting important applications such as dealing with longer histories in chat applications. To this end, sequence parallelism (SP) has emerged as a technique aimed at alleviating activation memory footprints during the training of Transformers. In SP, the input tensor of each Transformer layer is divided along the sequence dimension, allowing for parallel computation across multiple GPUs. This segmentation, in conjunction with activation recomputation, results in a substantial reduction in activation memory requirements by a factor of $s_{sp}$. In this paper, we classify existing SP approaches into 3 stages, which correspond to slicing the sequence into Norm and Dropout modules, Linear modules, and MHA module. When enabled cumulatively:

\textbf{SP-1: Norm and Dropout modules}. As shown in Figure \ref{fig_background} (b), Megatron-LM capitalizes on TP to parallelize the linear layers and MHA, which are the most time-consuming components during training. Simultaneously, it employs SP on Norm and Dropout modules, effectively reducing the activation memory of these layers by a factor of $s_{sp}$, where $s_{sp}=s_{tp}$. To maintain consistency in computational results, it integrates necessary communications, including \texttt{all-gather} and \texttt{reduce-scatter} to transfer the activation in forward and backward passes. When the activation size increases with the sequence length, this way of communicating the activation will incur a high overhead.

\textbf{SP-2: Add Linear modules}. DeepSpeed Ulysses utilizes sequence parallelism on Linear, Norm, and Dropout layers, as shown in Figure \ref{fig_background} (c). An \texttt{all-to-all} communication is applied to the output of $\emph{Linear}_\emph{qkv}$ with a shape of $B \times 3 \times S/s_{sp} \times H$. This enables each GPU to receive the complete sequence of $Q$, $K$, and $V$ (each with a shape of $B \times S \times H/s_{sp}$) for a subset of attention heads. Subsequently, FlashAttention is employed for highly efficient MHA computation with a constraint that the number of attention heads $D$ should not exceed $s_{sp}$. Another \texttt{all-to-all} communication is then used to gather the MHA layer results and the results are re-partitioned along the sequence dimension.  While DeepSpeed Ulysses achieves a reduction in activation memory usage and communication size per GPU by a factor of $s_{sp}$ compared to Megatron-LM, it requires the complete weight matrices of all Linear layers on each GPU. Therefore, DeepSpeed Ulysses applies additional techniques such as ZeRO to reduce the memory overhead of model states.
 
\textbf{SP-3: Add MHA modules}. Existing systems like \cite{ColosslaiSP, BPT2, li2023lightseq} apply sequence parallelism on all modules including MHA. In MHA, these approaches' division of $Q$, $K$, and $V$ along the sequence dimension is coupled with a series of peer-to-peer communication operations to facilitate the cooperative attention computation. This allows for a higher parallelism dimension than the number of attention heads ($s_{sp} > D$) and enables the training of models with significantly long sequences, reaching up to 1 million tokens. However, the division in MHA is incompatible with memory-efficient mechanisms like FlashAttention, which can reduce the memory complexity from $O(S^2)$ to $O(S)$. 

%% file: sections/searchspace.tex
\section{Search Space Analysis}
\label{sec:seach_space}

\begin{figure}[t]
    \centering
    \includegraphics[width=0.9\linewidth]{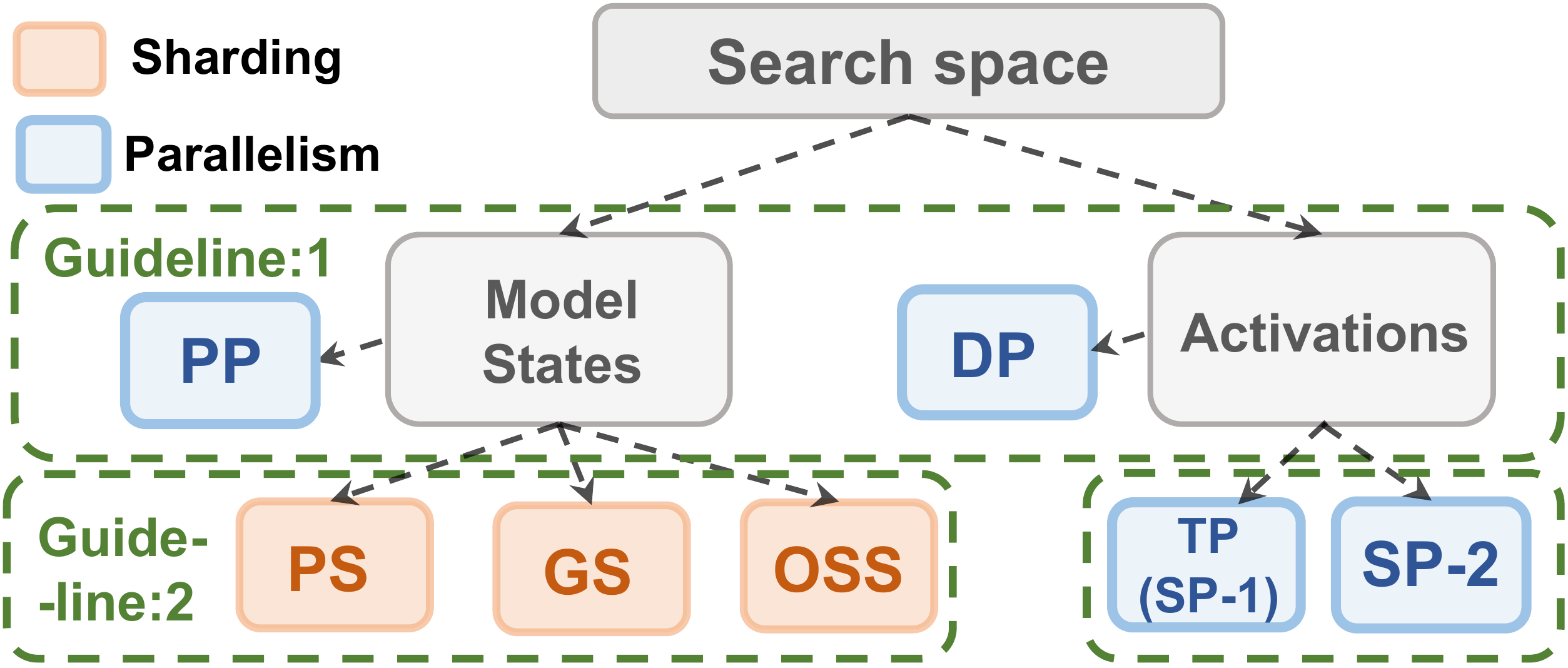}
    \caption{Search space overview. The orange and blue boxes are the dimensions in \sysname{}'s search space.}
    \label{Guidelines}
\end{figure}


\subsection{Search Space Overview}

\sysname{} is designed to efficiently handle the memory requirements of model states and activations, along with reducing the communication overhead and improving the training performance in distributed settings. 
Initially, we elaborate on the rationale behind the inclusion of specific strategies in the search space and justify their relevance.

\textbf{Guideline 1: Decouple the memory management strategy for model states and activations.} The memory consumption of large model training can be classified into two parts: model states (encompassing parameters, gradients, and optimizer states) and activation \cite{ZeRO}. As elucidated in Section~\ref{subsec_memory}, the memory allocation for model states represents a minor fraction of the overall memory in long-sequence training, with activation memory emerging as the primary bottleneck. 
The activation-reducing strategy SP-1 in Megatron-LM lacks the distinction between model states and activations during partitioning. 
This deficiency becomes apparent when attempting to fit activations within the memory constraints of a device, leading to a challenging dilemma. Simultaneously partitioning the sequence into smaller slices is hindered by the need to maintain constant sizes for model state slices, ensuring the communication cost of the model state remains constant.
To overcome this limitation, we propose an approach that separately partitions the model states and activations. 
This tailored partitioning allows for independent tuning of each aspect, presenting a promising avenue for more efficient memory and communication management, especially in the context of long-sequence giant model training.


\textbf{Guideline 2: Decouple memory management for model states into $P$, $G$, and $OS$.}
Despite the utilization of SP-2 by DeepSpeed, which employs ZeRO-3 to reduce the memory usage of model states by a factor of the world size $n$, there exists a limitation in considering the diverse memory footprints associated with the three components of model states—parameters ($P$), gradients ($G$), and optimizer states ($OS$), as outlined in Table \ref{tab:model_used_memory}. Based on this observation, we further decouple the memory management for model states into the management of $P$, $G$, and $OS$. This refined approach aims to better address the specific memory requirements of each component, potentially optimizing memory utilization in the context of long-sequence training.

\textbf{Guideline 3: Optimize communication by trading partial redundancy.}
A notable performance gap exists between intra- and inter-node network bandwidth and latency \cite{ZeRO++}, posing constraints on training efficiency.
In contrast to memory management for model states with zero redundancy, strategically grouping GPUs with slight memory redundancy for $P$, $G$, and $OS$ can effectively reduce communication costs. This reduction is achieved by reducing both the size and frequency of expensive inter-node communication. A relevant example is MiCS \cite{MiCS}, which introduces a cluster grouping strategy, partitioning all model states within each group and then replicating them across different groups.

Considering \textbf{Guidelines 1 $\sim$ 2}, we categorize GPU memory management dimensions into two main categories: model states and activations, as illustrated in Figure \ref{Guidelines}. Model state memory is governed by the pipeline parallelism (PP) strategy and three sharding strategies: parameter sharding (PS), gradient sharding (GS), and optimizer states sharding (OSS). Activation memory, on the other hand, is managed by data parallelism (DP), tensor parallelism (TP), and sequence parallelism (including SP-1 and SP-2).
Overall, \sysname{} define 4 parallel dimensions and 3 sharding dimensions as shown in the blue and orange blocks of Figure \ref{Guidelines}, and presented as $\mathbb{S}= [b,s,a,s_{pp},s_{dp},s_{tp},s_{sp},s_{ps},s_{gs},s_{oss}]$.  Based on \textbf{Guideline 3}, different values of $s_{i}$ exert a significant influence on communication costs.
We find that the expensive inter-node communication of certain $i$ is could be reduced by optimizing $s_{i}$ to achieve partially redundant storage.
This strategic adjustment enables the consolidation of communication for $i$ within the same node with higher bandwidth. 
Next, we introduce how \sysname{} models the memory and communication cost of model states and activations for each item in $\mathbb{S}$.

\subsection{Memory \& Communication Analysis}
\label{subsec:memoryAnalysis}

We explore the memory footprint and communication costs of model training, considering different combinations of parallelism and sharding strategies. Table \ref{tab:transformer-parameters} shows notations used in this analysis. Our focus here is the costs of the $L$ Transformer layers of a model. The embedding layer and head layer of the model will be discussed in the subsequent section.

\sysname{} optimizes the memory usage by storing activations at Transformer layer boundaries and recomputing the remaining activations in the backward pass. When $a=1$, \sysname{} requires the storage of $2bSH$ bytes for activations of a Transformer layer, applicable in scenarios without pipeline or sequence parallelism. Conversely, if $a=1$, \sysname{} utilizes $34bSH$ bytes for activation storage. Furthermore, to enhance the efficiency of multi-head attention computation, \sysname{} uses the fused kernel of FlashAttention by default, which already incorporates the \emph{selective activation recomputation} technique to minimize the memory usage during attention computation \cite{flashatten1}. 
The total GPU memory consumed by activations across all $L$ Transformer layers is given by:
\begin{equation*}
M_{{ACT}}(s_{sp}=1,s_{pp}=1) = \sum_{i=0}^{L-1}(2bSH + 32bSH(1-a)).
\end{equation*}

\subsubsection{Basic Communication Cost Model}
\label{subsec:basic_communication_cost_model}

\begin{figure}[t]
    \centering
    \includegraphics[width=\linewidth]{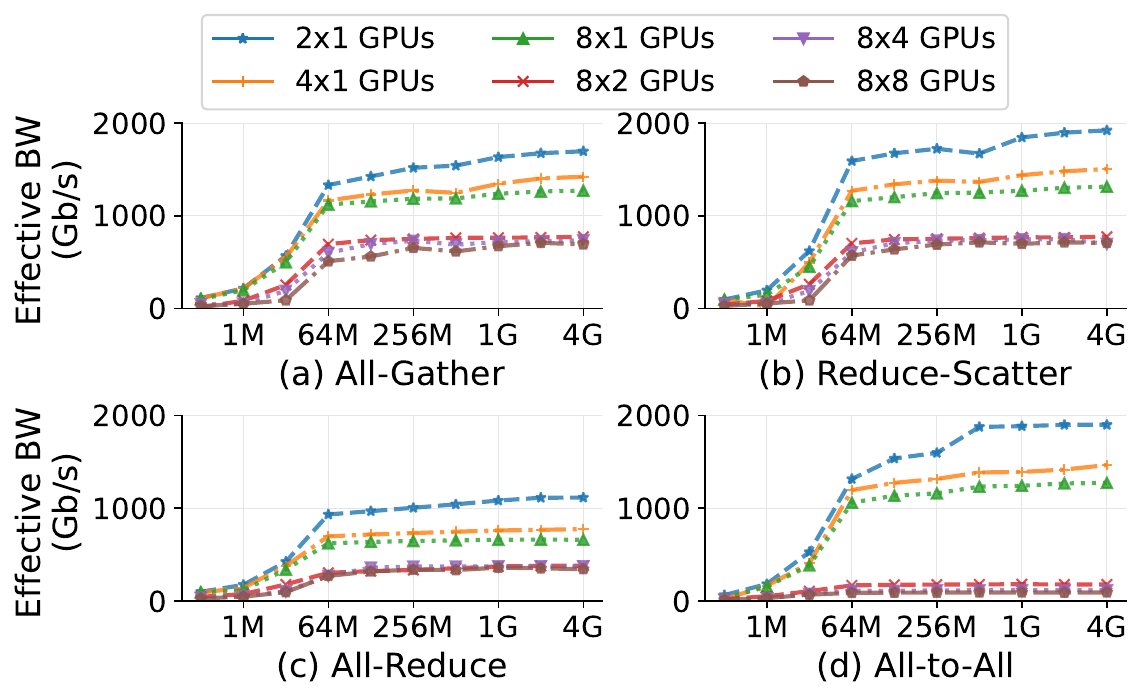}
    \caption{Performance evaluation of collective communication operations using NCCL on Nvdia Ampere GPUs. For configurations with up to 8 GPUs, the tests were executed within a single node. For scenarios with more than 8 GPUs, we extended the evaluation across nodes interconnected by 4x Mellanox Infiniband HDR NICs.}
    \label{fig:comm}
\end{figure}

\sysname{} uses a device mesh to represent the communication pattern \cite{Alpa,Unity,OneFlow}. 
This logical view, a 2-dimensional abstraction of physical devices, illustrates their ability to communicate along the first and second dimensions, each with distinct bandwidths. To specify the communication characteristics of participant GPUs ($p$), we employ a superscript notation. Specifically, $p^0$ denotes participant GPUs communicating along the 0-th axis via intra-node networks like NVLink, while $p^1$ signifies communication along the 1-th axis through inter-node networks like Infiniband.
 \sysname{} uses the cost estimation formula $\tau(o, v, p^{\{0,1\}}) = v/w(o,v,p^{\{0,1\}})$ to assess the time consumption of collective communication operator ($o$) with a given data size ($v$) and a specified participant GPU count ($p$). Here, $w(o,v,p^{\{0,1\}})$ represents the effective bandwidth obtained through performance profiling on the target GPU cluster, as illustrated in Figure \ref{fig:comm}. Our focus in this study revolves around four key collective communication operations: \texttt{all-reduce}, \texttt{all-gather}, \texttt{reduce-scatter}, and \texttt{all-to-all}. \sysname{} has the flexibility to extend its cost estimation to other operators as needed.




\subsubsection{Analysis of Parallelism Strategies}

We conduct the analysis of the memory footprint and communication cost when \sysname{} adjusts parallelism strategies using $s_{dp}$, $s_{pp}$, $s_{tp}$, and $s_{sp}$.

\textbf{Data Parallelism \& Gradient Accumulation.} \sysname{}  uses data parallelism to distribute data samples of a single step across $s_{dp}$ GPU worker groups, with each group maintaining a replica of the model. Additionally, \sysname{} facilitates the emulation of a larger batch size by cumulatively aggregating gradients from $n$ micro-batches before executing the weight update. 
These combined strategies prove beneficial in alleviating the activation memory footprint $M_{{ACT}}$, which escalates linearly with the micro-batch size $b$. It is important to note that data parallelism introduces a necessary communication cost for synchronizing gradients, and we will analyze such costs in the context of the optimizer states sharding strategy.

\textbf{Pipeline Parallelism.} \sysname{} can harness pipeline parallelism to evenly partition $L$ Transformer layers into $s_{pp}$ stages and distribute them across GPUs, effectively reducing the memory footprint of $P$, $G$, and $OS$ by a factor of $s_{pp}$. By default, \sysname{} employs the 1F1B scheduler, which divides an input batch into $n$ micro-batches and processes the forward and backward computations alternately \cite{GPipe,PipeDream}. Consequently, the first stage must store activations for as many micro-batches as the degree of pipeline parallelism, while subsequent stages gradually store fewer activations \cite{PipeDream}. In case of $s_{pp}>1$, we define the maximum GPU memory usage as follows:
\begin{align*}
& M_{{P,G,OS}}(s_{pp} \geq 1) = {M_{P,G,OS}(s_{pp}=1)}/{s_{pp}}, \\
& M_{{ACT}}(s_{pp} \geq 1) = \min(s_{pp},n) \times {M_{ACT}(s_{pp}=1)}/{s_{pp}}.
\end{align*}
In the pipeline parallelism setup, consecutive stages exchange intermediate data through point-to-point (P2P) communication. We exclude the consideration of such communication costs in this work for two reasons: 1) P2P communications occur between stages rather than within each Transformer layer, resulting in a much smaller communication size compared to other aspects; 2) such communication can be overlapped with computation during model training \cite{zhuang2023optimizing}. Additionally, each pipeline stage exhibits data dependencies on others, leading to computation stalls, commonly known as a "pipeline \emph{bubble}," which will be discussed in the subsequent section.

\textbf{Tensor Parallelism.} When $s_{tp}>1$, \sysname{} incorporates tensor parallelism to parallelize the computationally-intensive linear layers and Multi-Head Attention. Simultaneously, it defaults to utilizing sequence parallelism for normalization and dropout modules, resulting in $s_{sp}=s_{tp}>1$. In this scenario, \sysname{} effectively reduces the GPU memory usage of $P$, $G$, $OS$, and $ACT$ by a factor of $s_{tp}$, as expressed by the equation:
\begin{equation*}
M_{{P,G,OS,ACT}}(s_{tp} \geq 1) = {M_{P,G,OS,ACT}(s_{sp}=s_{tp}=1)}/{s_{tp}}.
\end{equation*}
As shown in Figure \ref{fig_background}, \sysname{} uses one \texttt{all-gather} and one \texttt{reduce-scatter} operation on each Transformer layer per micro-batch during the forward pass when $s_{sp}=s_{tp}>1$. In the backward pass, the insertion of two \texttt{all-gather} and one \texttt{reduce-scatter} communication operations occur per micro-batch. The communication time for the $i$-th layer of a step, introduced by tensor parallelism, is as follows:
\begin{align*}
T^{i}_{comm}(s_{tp}=s_{sp} > 1) = 
& 2n\tau(\texttt{reduce-scatter}, 2bSH, s_{tp}) + \\
& 3n\tau(\texttt{all-gather}, 2bSH, s_{tp}).
\end{align*}

\textbf{Sequence Parallelism.}  \sysname{} employs sequence parallelism for all modules except MHA when $(s_{sp}>1, s_{tp}=1)$. As illustrated in Figure \ref{fig_background} (c), \sysname{} re-partitions through \texttt{all-to-all} before and after Multi-Head Attention, ensuring the hidden dimension of this module slice aligns with the sequence, similar to other modules. Consequently, \sysname{} reduces the memory footprint of ACT by a factor of $s_{sp}$, as indicated by the equation:
\begin{align*}
M_{{ACT}}(s_{sp} \geq 1, s_{tp}=1) = {M_{ACT}(s_{sp}=s_{tp}=1)}/{s_{sp}}.
\end{align*}
In each Transformer layer during both forward and backward passes, \sysname{} incorporates two \texttt{all-to-all} communication operations per micro-batch. The initial \texttt{all-to-all}  exchanges $6bSH$ bytes of data, followed by a second exchange of $2bSH$ bytes of data. The communication time attributable to sequence parallelism for the $i$-th layer of a step is:
\begin{align*}
T^{i}_{comm}(s_{sp} > 1, s_{tp}=1) =
& 2n\tau(\texttt{all\_to\_all}, 6bSH, s_{sp}) + \\
& 2n\tau(\texttt{all\_to\_all}, 2bSH, s_{sp}).
\end{align*}
It is noteworthy that this strategy necessitates the complete weight matrices on each GPU for computation. 

\subsubsection{Analysis of Sharding Strategies}

\sysname{} uses a mix of parallelism strategies to distribute Transformer layers and modules efficiently across GPUs. Parameters, gradients, and optimizer states may be replicated or sharded among GPUs based on two configurations: 
\begin{itemize}[topsep=0pt]
\setlength{\itemsep}{0pt}
\setlength{\parskip}{0pt}
\setlength{\parsep}{0pt}
\item  {Config-1: $s_{dp} \geq 1, s_{pp} \geq 1, s_{tp}=s_{sp} > 1$.}
GPUs with the same tensor and pipeline parallelism rank hold replicated parameters, ensuring synchronized training. 
\item {Config-2: $s_{dp} \geq 1, s_{pp} \geq 1, s_{tp}=1, s_{sp} \geq 1$.} GPUs within the same pipeline stage jointly hold replicated parameters along with their gradients and optimizer states.
\end{itemize}

\begin{figure}[t]
    \centering
    \includegraphics[width=\linewidth]{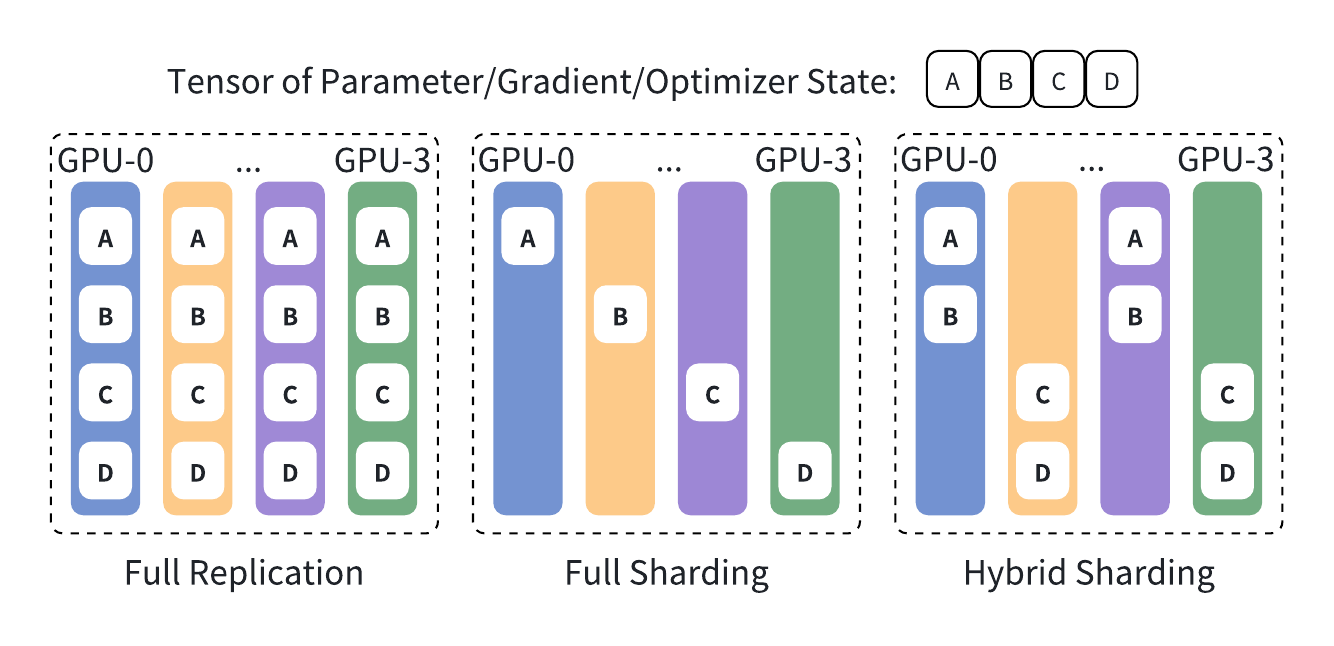}
    \caption{Three Sharding strategies.}
    \label{Fig:Sharding}
\end{figure}

As illustrated in Figure \ref{Fig:Sharding}, \sysname{} incorporates three sharding strategies, impacting memory footprint and communication cost. The sharding factor $F$ denotes the number of GPUs over which the tensor is sharded. For a tensor with $E$ elements and $R$ GPUs, the three strategies could be outlined as follows: 
\begin{itemize}[topsep=0pt]
\setlength{\itemsep}{0pt}
\setlength{\parskip}{0pt}
\setlength{\parsep}{0pt}
\item {Full Replication ($F=1$):} A complete replica of the tensor with $E$ elements is stored on each of the $R$ GPUs.
\item {Full Sharding ($F=R$):}
Each GPU stores only $E/R$ elements, and the full tensor can be reconstructed by performing an \texttt{all-gather} across all $G$ GPUs.
\item {Hybrid Sharding ($1<F<R$):}
This strategy combines sharding and replication. Each GPU holds $E/F$ elements, and each element is replicated on $R/F$ GPUs. The full tensor can be reconstructed by employing an \texttt{all-gather} across the $F$ GPUs.
\end{itemize}

Unlike ZeRO-1, ZeRO-2 \cite{ZeRO}, MiCS \cite{MiCS}, ZeRO-3 \cite{PyTorchFSDP} and ZeRO++ \cite{ZeRO++}, which use static sharding strategies for parameters, gradients, and optimizer states, \sysname{} takes a more flexible approach: it applies three separate sharding strategies — determined by distinct factors $s_{ps}$, $s_{gs}$, and $s_{oss}$ — to parameters, gradients, and optimizer states. This flexibility allows for a balance between communication and memory, offering tailored performance optimization during training.

\begin{figure}[t]
    \centering
    \includegraphics[width=\linewidth]{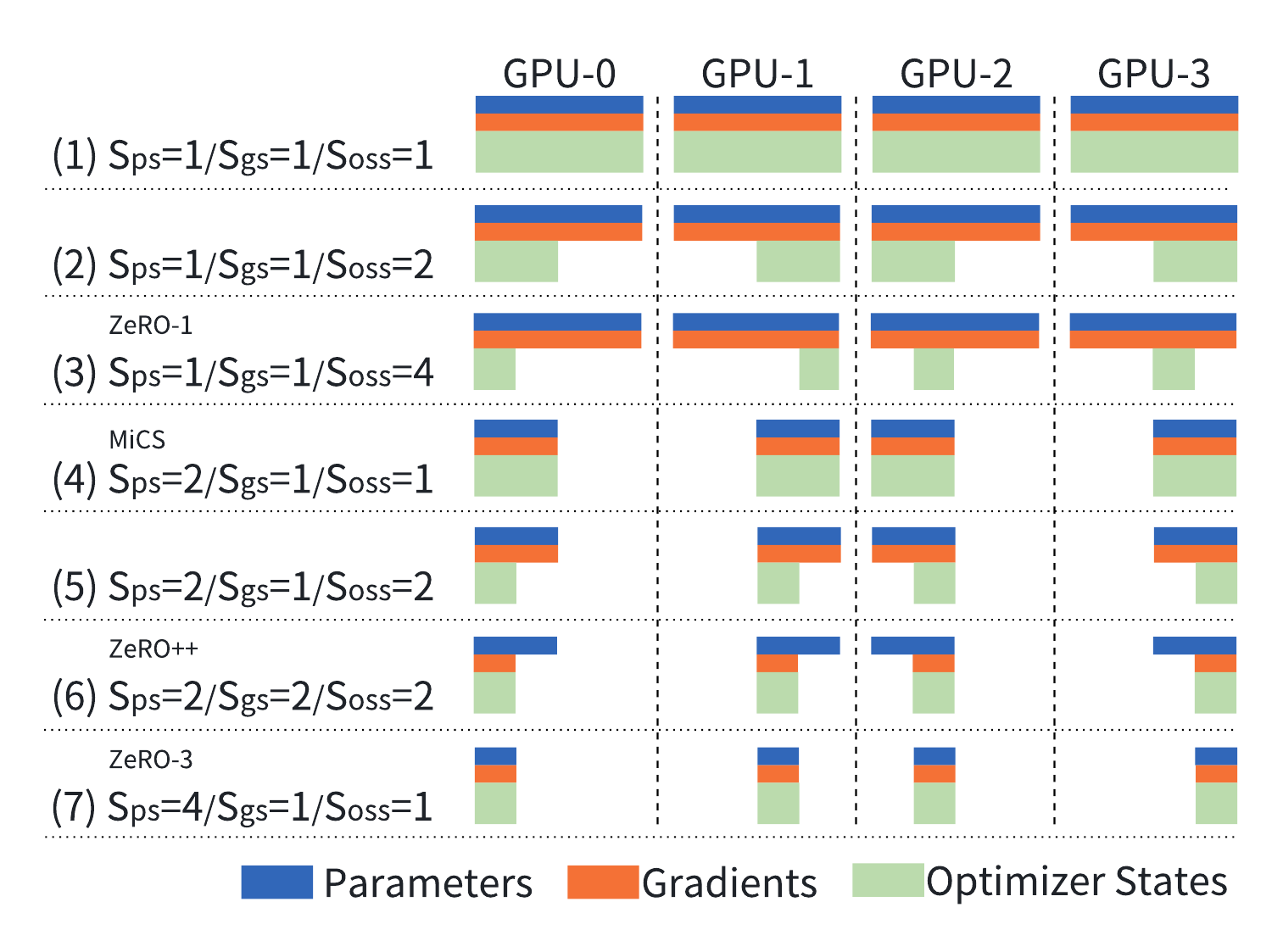}
    \caption{Overview of parameter sharding, gradient sharding, and optimizer state sharding with various sharding factors.}
    \label{Fig:Sharding_detail}
\end{figure}

\textbf{Parameter Sharding.} When parameters are replicated across GPUs, \sysname{} offers the capability to shard these parameters with a sharding factor $s_{ps}$. This  strategy, illustrated in Figure \ref{Fig:Sharding_detail}, effectively reduces GPU memory footprint of parameters by a factor of $s_{ps}$, as expressed by the equation:
\begin{align*}
M_{P}(s_{ps} \geq 1) = {M_{P}(s_{ps}=1)}/{s_{ps}}.
\end{align*}
During both the forward and backward passes of each micro-batch, \sysname{} orchestrates the collection of parameter shards from other GPUs to reconstruct the complete set of model weights necessary for computations. In the forward pass, a single \texttt{all-gather} operation on the corresponding parameter sharding group is employed. Subsequently, during the backward pass, each GPU computes gradients for the entire set of model parameters collected by another \texttt{all-gather} operation. Following this, \sysname{} utilizes the \texttt{reduce-scatter} function to aggregate and distribute gradients across the parameter sharding group.  The communication time attributable to parameter sharding for the i-th layer of a step is:
\begin{align*}
T^{i}_{comm}(s_{ps} > 1) = 
& 2n\tau(\texttt{all-gather}, {2\Psi}/{s_{tp}}, s_{ps}) + \\
& n\tau(\texttt{reduce-scatter}, {2\Psi}/{s_{tp}}, s_{ps}),
\end{align*}
where ${\Psi}/{s_{tp}}$ is the parameter count of a Transformer layer partitioned by tensor parallelism. Additionally, the sharding factor $s_{ps}$ is constrained by:
\begin{equation*}
s_{ps} \in [1, {N}/({s_{pp} \cdot s_{tp}})] \cap \mathbb{Z}.
\end{equation*}
This constraint ensures the feasibility of parameter sharding, considering various configurations. For example, when $s_{tp}=s_{sp}>1$, ${N}/({s_{pp} \cdot s_{tp}})$ equates to $s_{dp}$. When $s_{tp}=1$ and $s_{sp}>1$, the maximum of $s_{ps}$ could be higher than $s_{dp}$, allowing for more flexibility in optimizing $s_{ps}$.

\textbf{Optimizer State Sharding.} 
When parameters are replicated across $R$ GPUs and subjected to the parameter sharding operation with a factor $s_{ps}$, a parameter may still be distributed among multiple GPUs. For instance, in Figure \ref{Fig:Sharding_detail}(6), with $s_{ps}=2$, GPU-0 and GPU-2 store the same parameters. Following the MiCS approach \cite{MiCS}, the gradients and optimizer states associated with these parameters are also replicated on both GPU-0 and GPU-1.
In contrast, \sysname{} offers enhanced flexibility by allowing the sharding of gradients and optimizer states independently using factors $s_{gs}$ and $s_{oss}$. This decoupling of sharding factors for parameters, gradients, and optimizer states provides finer control over the distribution strategy, enabling more efficient memory utilization and communication patterns during training. 
When $s_{oss}>1$, the memory footprint of optimizer states can be further reduced, as expressed by the following equation:
\begin{align*}
M_{{OS}}(s_{oss} \geq 1) = M_{OS}(s_{ps}=1,s_{oss}=1)/(s_{ps} \cdot s_{oss}),
\end{align*}
subject to the constraint:
\begin{equation*}
s_{oss} \in [1, {N}/({s_{pp} \cdot s_{tp} \cdot s_{ps}})] \cap \mathbb{Z}.
\end{equation*}
This formulation ensures that the sharding factor $s_{oss}$ remains within a valid range, aligning with the overall parallelism and sharding strategy for optimizer states.

After the last micro-batch's backward pass, each GPU is responsible for updating the weights of parameters based on the optimizer states. Prior to the parameter update stage, each GPU is required to collect and aggregate gradients for parameters within its optimizer states. To streamline this process, \sysname{} employs one \texttt{all-reduce} operation on gradients across GPUs that share the same set of parameters. For instance, in Figure \ref{Fig:Sharding_detail}(2), \sysname{} conducts \texttt{all-reduce} on all four GPUs to synchronize the gradients. Subsequently, GPU-0 and GPU-2 select the first half of the aggregated gradients for weight updates. In Figure \ref{Fig:Sharding_detail}(6), \sysname{} conducts \texttt{all-reduce} specifically on GPU-0 and GPU-2, obtaining the aggregated gradients necessary for weight updates. 
Following the optimizer update stage, \sysname{} employs one \texttt{all-gather} operation on updated parameters across GPUs within the same optimizer state sharding group. This ensures that each GPU receives all the updated values for parameters stored in its local memory. For instance, in Figure \ref{Fig:Sharding_detail}(2), \sysname{} conducts \texttt{all-gather} on GPU-0 and GPU-1.
The communication time associated with optimizer states for the $i$-th layer is articulated as follows:
\begin{align*}
T^{i}_{comm}(s_{oss} \geq 1) =  
\tau(\texttt{reduce-scatter}, {2\Psi}/({s_{tp} \cdot {s_{ps}}}), s_{oss}) \\
+ \tau(\texttt{all-reduce}, {2\Psi}/({s_{tp} \cdot {s_{ps}}}), {N}/({s_{pp} \cdot s_{tp} \cdot s_{ps}})).
\end{align*}

\noindent\textbf{Gradient Sharding.} 
When $s_{ps} > 1$, the gradient is inherently sharded with parameters. If there are still replicas of the gradient in the cluster, \sysname{} support gradient sharding with factor $s_{gs}$. This is done to effectively mitigate memory usage for gradients, particularly in the context of gradient accumulation. The memory footprint of gradients is expressed as:
\begin{align*}
M_{G}(s_{gs} \geq 1) = M_{G}(s_{ps}=1, s_{gs}=1)/(s_{ps} \cdot s_{gs}).
\end{align*}
For simplicity, we restrict the choice of $s_{gs}$ in this work: 
\begin{align*}
s_{gs} \in \{1, s_{oss}\}. 
\end{align*}
For example, in Figure \ref{Fig:Sharding_detail}(6), gradients are sharded with $s_{gs}=2$. 
To facilitate gradient accumulation, \sysname{} operates \texttt{all-reduce} on each micro-batch, excluding the last one, to aggregate gradients across GPUs with the same portion of parameters. The associated communication cost is given by:
\begin{align*}
\small
& T^{i}_{comm}(s_{gs}\!>\!1)\!=\!(n\!-\!1) \tau(\texttt{all-reduce}, \frac{2\Psi}{s_{tp}\!\cdot\!{s_{ps}}}, \frac{N}{s_{pp}\!\cdot\!s_{tp}\!\cdot\!s_{ps}}).
\end{align*}

\subsubsection{Summary}

In the training of transformer-based models with a combination of parallelism and sharding strategies, we provide a comprehensive analysis of communication and memory considerations. Specifically, the GPU memory usage of all transformer layers, denoted as $M_{tf}$, can be articulated as follows:
\begin{equation}
\small
\label{eq:memoryconstrain}
M_{tf} = \underbrace{\frac{2L\Psi}{s_{pp} \cdot s_{tp}}(\frac{1}{s_{ps}}+\frac{1}{s_{gs}}+\frac{1}{s_{oss}})}_{\text{Memory used by P, G and OS}} + \underbrace{\frac{bSHL (34-32a))}{s_{pp} \cdot s_{sp}}}_{\text{Memory used by ACT}}.
\end{equation}
Note that we focus on the scenario where the pipeline parallelism size exceeds the micro-batch number here ($s_{pp} > n$). Additionally, the communication time of $i$-th layer during a step, denoted as $T_{comm}^i$, can be expressed as:
\begin{equation}
\small
\label{eq:total comm}
T_{comm}^i = \underbrace{T^i_{comm}(s_{tp},s_{sp})}_{\text{Communicate ACT}}+\underbrace{T^i_{comm}(s_{ps},s_{gs},s_{oss})}_{\text{Communicate P, G}}.
\end{equation}
Communication may be overlapped with computation, which will be explored in subsequent discussions.

%% file: sections/design.tex
\section{System Design}

In this section, we delve into how \sysname{} navigates the search space, which comprises micro-batch size, micro-batch number, activation recomputation status, four parallelism strategies and three sharding strategies, denoted as $\mathbb{S}= [b, n, a, s_{pp},s_{dp},s_{tp},s_{sp},s_{ps},s_{gs},s_{oss}]$. We focus on how \sysname{} systematically optimizes the overall training performance within this defined space.

\subsection{System Architecture \& WorkFlow}

\begin{figure}[t]
    \centering
    \includegraphics[width=1\linewidth]{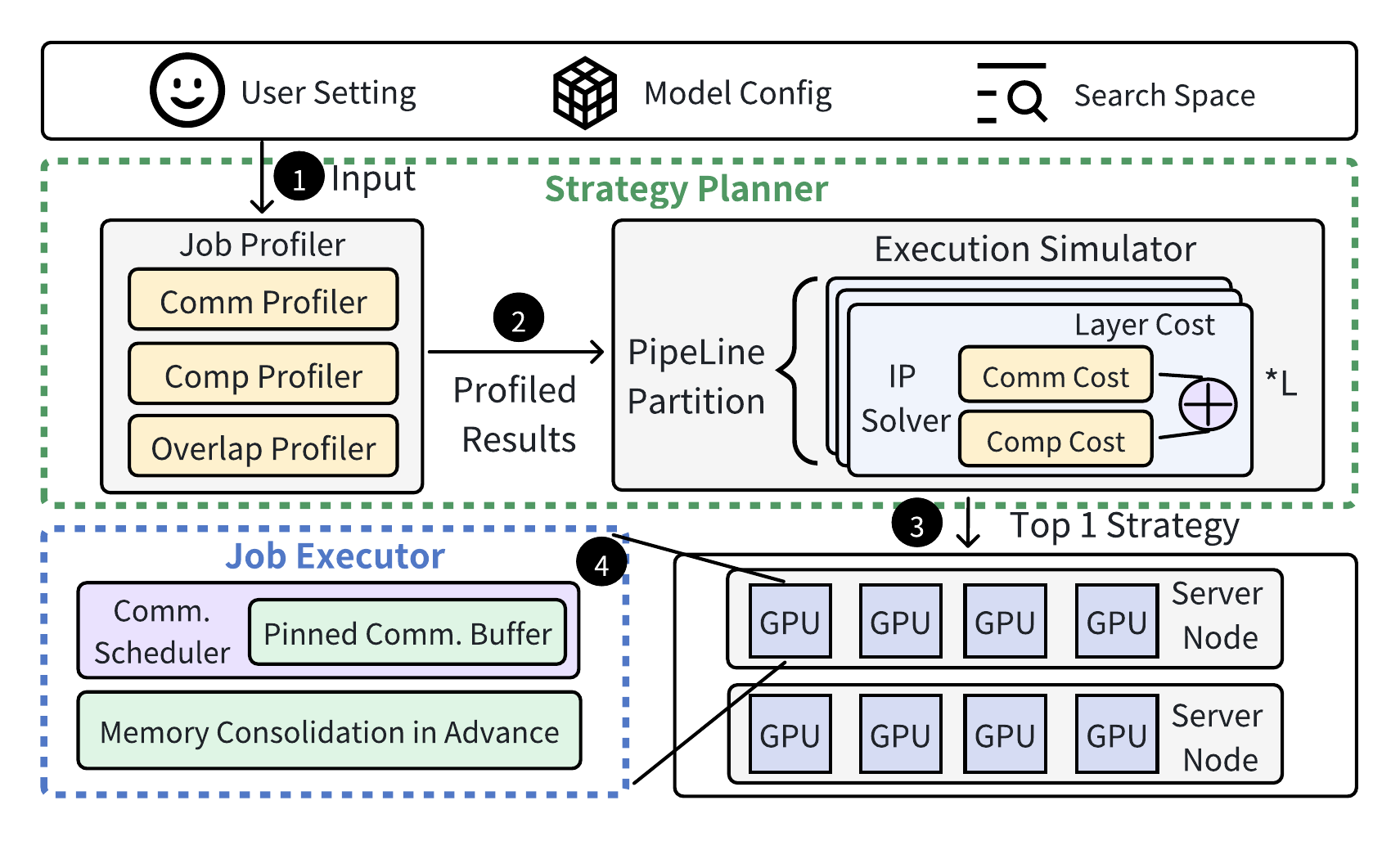}
    \caption{Overview of \sysname{} architecture and workflow.}
    \label{fig:framework}
\end{figure}

Illustrated in Figure~\ref{fig:framework}, \sysname{} comprises two key components: Strategy Planner and Job Executor.
(1) The Strategy Planner is dedicated to identifying the optimal strategy $\mathbb{S}$ for parallelizing model training. This involves two core modules: firstly, the Job Profiler, which meticulously profiles the submitted workload to extract crucial execution information; and secondly, the Execution Simulator, tasked with simulating various strategies within $\mathbb{S}$ to pinpoint the most effective solution.
(2) The Job Executor is responsible for the actual execution of the training job with the given training strategy $\mathbb{S}$ on GPU nodes.
This component integrates two essential modules to enhance training performance. The Communication Scheduler, through its ability to overlap computation and communication at a fine granularity, significantly mitigates communication overhead. Simultaneously, the Fragmentation Memory Manager ensures unified GPU memory management, effectively reducing the occurrence of OOM errors attributed to GPU memory fragmentation, particularly in the context of long-sequence training scenarios.

\textbf{Workflow.} 
\circled{1}  \sysname{}  commences with users specifying transformer model architecture like attention head number and layer number, hyper-parameters such as sequence length and global batch size, and training cluster settings like total GPU number and memory capacity.  
\circled{2} The Job Profiler conducts a comprehensive analysis by profiling the effective bandwidth of used collective communication operations, computation times for modules within the model, and identifying instances of communication-computation overlap (Section~\ref{sec:profiler}). The gained profiling results drive the Execution Simulator to estimate the memory cost and step time for various combinations of parallelism and sharding strategies (Section~\ref{sec:execution_simulator}). Leveraging an optimization problem solver, \sysname{} identifies the \emph{top-10} strategies, and real system profiling is employed to determine the best one \circled{3}. This exploration leads to the identification of a parallelization strategy with the shortest step time, ensuring the absence of out-of-memory errors.
\circled{4} Then, the Job Executor runs the training job utilizing the corresponding strategy with high training speed and memory utilization.

\subsection{Job Profiler}
The Job Profiler consists of three parts: (1)\texttt{VPro} profiles the effective bandwidths for NCCL communication offline. (2) \texttt{UPro} conducts detailed profiling of a single forward pass of a transformer layer online. 
(3) \texttt{OPro} profiles the overlap between computation and communication offline. Details of the three components are as follows:

\label{sec:profiler}
\textbf{Communication Profiler (\texttt{VPro})}. 
For an execution plan $\mathbb{S} = [b, n, a, s_{pp},s_{dp},s_{tp},s_{sp},s_{ps},s_{gs},s_{oss}]$, \sysname{} leverages real-system profiling to determine the effective bandwidth of four used collective communication operations (i.e. \texttt{all-reduce}, \texttt{all-gather}, \texttt{reduce-scatter}, and \texttt{all-to-all}) across various communication sizes and device meshes. Figure \ref{fig:comm} gives an example of the profiling results. Thus, \sysname{} estimates the communication latency induced by parallelism and sharding strategies according to Equation \ref{eq:total comm}: $T_{comm}^i=\texttt{VPro}(\mathbb{S})$. 



\textbf{Computation Profiler (\texttt{UPro})}.
\sysname{} conducts computation time profiling for $i$-th transformer layer with the specified execution plan $\mathbb{S}$: $T_{comp}^i=\texttt{UPro}(\mathbb{S})$.  As discussed in Section \ref{MemoryFootprint}, \sysname{} identifies the self-attention and linear modules as the key contributors to computation time in the Attention and MLP block.
For the self-attention module, \sysname{} profiles the execution time of \texttt{self-atten($b,S,H,D/s_{sp}$)} on a single GPU using FlashAttention.
Regarding the linear modules, if $s_{sp}>1$ and $s_{tp}=1$, \sysname{} profiles the execution time of applying the input tensor with dimension $(b,S/s_{sp},H)$ on the {Linear} module. In the scenario where $s_{tp}>1$, \sysname{} profiles the computation time of each linear module based on its row-partitioning or column-partition strategy, as discussed in Section \ref{sec:seach_space}. 


\textbf{Overlap Profiler (\texttt{OPro})}. 
\sysname{} supports the concurrent execution of communication operations alongside computation kernels. Notably, in the forward pass, it enables the simultaneous execution of computation and the \texttt{all-gather} communication resulting from parameter sharding. In the backward computation phase, \sysname{} extends this capability to various communication tasks, such as \texttt{all-gather} and \texttt{reduce-scatter} induced by tensor parallelism and parameter sharding, as well as \texttt{all-reduce} from gradient sharding and optimizer states sharding. 
Despite the widespread practice of estimating total training time by taking the maximum value between profiled communication time (\texttt{VPro}) and computation time (\texttt{UPro}), this approach may lack accuracy. Communication primitives, such as NCCL \cite{Allreduce}, introduce a slowdown in both computation and communication due to resource contention in GPU streaming multiprocessors \cite{Galvatron}. This contention varies with model sizes and sequence lengths, leading to performance degradation. \sysname{} defines a slow down ratio $\mathcal{R}$ for \texttt{OPro} to tackle this challenge:
$$\texttt{OPro}(\mathbb{S}) = \mathcal{R} \cdot \max (\texttt{VPro}(\mathbb{S}),\texttt{UPro}(\mathbb{S})).$$
In this study, the parameter $\mathcal{R}$ varies within the range of $1.25$ to $1.35$ as determined through evaluations.




\subsection{Communication Scheduler}
The Communication Scheduler module in \sysname{} strategically orchestrates communication and computation to enhance overall system performance. Specifically, our approach focuses on optimizing the overlap during the backward pass by implementing a refined intra-layer strategy. This involves leveraging parameter pre-fetching (\texttt{all-gather}) and gradient synchronization (\texttt{reduce-scatter}) at the granularity of modules, such as the linear module.
In the forward pass, we employ an inter-layer overlap strategy to mitigate latency by pre-fetching the entire set of parameters for upcoming layers while concurrently computing the current layer. This is particularly beneficial when $s_{ps} > 1$. The iterative selective overlap approach adeptly manages both communication overhead and computation execution time, resulting in a noteworthy enhancement in overall system performance. For more design and implementation detail refers to Appendix \ref{sec:overlap}.

\subsection{Memory Fragmentation Manager}

The \texttt{cudaMalloc} and \texttt{cudaFree} interfaces cause substantial latency and impose significant synchronization overhead across CUDA streams. To mitigate these challenges, memory pools, such as CNMeM \cite{CNMeM} and the PyTorch memory allocator \cite{paszke2019pytorch}, serve as an intermediary layer to handle memory allocation and deallocation requests during model training. These memory pools proactively pre-allocate contiguous memory chunks of varying sizes and employ strategies like merging or splitting chunks to meet diverse memory requirements.

Unfortunately, scenarios involving frequent (de)allocation requests can lead to memory fragmentation, posing challenges to the efficiency of memory pools and potentially impacting the training of models during peak memory usage. This issue becomes more pronounced in the context of long sequence training with communication/computation overlapping, as this case necessitates the temporary allocation of large contiguous memory chunks. In response to this challenge, we propose two measures:  \textit{Pinned Communication Buffer} and  \textit{Memory Consolidation in Advance}. Specifically, the pinned communication buffer introduces dedicated memory pools for \texttt{all-gather} communications, implementing double buffer rotation to efficiently manage memory during forward and backward passes. The memory consolidation approach proactively consolidates small memory chunks and addresses challenges posed by fragmented memory space. For more design and implementation detail refers to Appendix \ref{sec:appendix_fragmentation}.

\subsection{Execution Simulator}
\label{sec:execution_simulator}

The execution simulator estimates training performance using a given execution plan $\mathbb{S}$ and obtains the best solution candidates by solving an integer programming problem.

We formulate $T_{fwd\_bwd}(\mathbb{S})$, representing the total time for all forward and backward passes in a training step with the given execution plan. The equation is given by:
\begin{equation*}
\small
  \label{eq:time_fwd_bwd}
  T_{fwd\_bwd}(\mathbb{S}) = (n+s_{pp}-1) \cdot (T_{other}(\mathbb{S}) + \lceil {L}/{s_{pp}} \rceil \cdot \texttt{OPro}(\mathbb{S})),
\end{equation*}
Here, $T_{other}$ is the execution time of other layers, such as the head layer and the embedding layer, which can be estimated through profiling. The execution time of a single micro-batch is given by $T_{other}(\mathbb{S}) + \lceil {L}/{s_{pp}} \rceil \cdot \texttt{OPro}(\mathbb{S})$. In the presence of a pipeline \emph{bubble} caused by data dependencies, the total time of all forward and backward passes is obtained by multiplying the single micro-batch time with $(n+s_{pp}-1)$.

Building upon equation \ref{eq:memoryconstrain}, which defines the GPU memory usage of a transformer layer, the total GPU memory usage $M(\mathbb{S})$ can be expressed as:
\begin{equation*}
\small
    \label{eq:time_memory}
    M(\mathbb{S}) = M_{tf}(\mathbb{S}) + M_{other}(\mathbb{S}),
\end{equation*}
where $M_{other}$ represents the memory used by head and embedding layers, as well as temporary buffers for communication and computation.

\sysname{} formulates an optimization problem to search for optimal execution plans by minimizing the sum of the execution time of forward-backward passes and the parameter update time, subject to various constraints. The integer programming problem is defined as follows:
\begin{align}
\label{eq:ipproblem}
\text{Minimize} \quad & T_{fwd\_bwd}(\mathbb{S}) + T_{update}(\mathbb{S}) \\
\text{Subject to} \quad 
& M(\mathbb{S}) \leq \texttt{GPU\_Memory\_Capacity} \label{eq:maxmemconstraint} \\
& b \cdot S \cdot n \cdot s_{dp} = B \label{eq:batchsizeconstraint} \\
& s_{dp} \cdot s_{sp} \cdot s_{pp} = N \label{eq:gpuconstraint} \\
& s_{ps} \cdot s_{pp} \cdot s_{tp} \cdot i = N \quad i \in \mathbb{Z} \\
& s_{ps} \cdot s_{oss} \cdot s_{pp} \cdot s_{tp} \cdot j = N \quad j \in \mathbb{Z} \\
& s_{gs} \in \{1, s_{oss}\}
\end{align}
This problem optimizes the execution plan with respect to GPU memory capacity (Equation \ref{eq:maxmemconstraint}), global batch size (Equation \ref{eq:batchsizeconstraint}), global GPU number (Equation \ref{eq:gpuconstraint}), and other considerations outlined in Section \ref{sec:seach_space}.

For each $s$ in $\mathbb{S}$, allocating specific tensors on the device mesh introduces multiple possible configurations (\ref{subsec:basic_communication_cost_model}), influencing the communication latency. Instead of exhaustively iterating through all potential GPU allocations for $\mathbb{S}$, we leverage practical insights to streamline the exploration of assignment strategies more efficiently.

\underline{Takeaway 1:} In cases where $s_{sp} \geq 1, s_{pp} \geq 1, s_{dp} \geq 1$, prioritize allocating ranks of the same tensor/sequence parallelism group with the fewest physical nodes. When activation memory usage significantly exceeds model state memory, communication to activation incurs substantial overhead. Therefore, give preference to utilizing high-performance intra-node bandwidth for activation communication.

\underline{Takeaway 2:} If $s_{ps} \geq 1, s_{gs} \geq 1, s_{oss} \geq 1$, prioritize allocating ranks of the same parameter sharding group with the fewest physical nodes. When $s_{ps}>1$, frequent \texttt{all-gather} communication can impact overall training performance. Therefore, favor the use of high-performance intra-node bandwidth for parameter pre-fetching.

Following the outlined takeaways, with a given execution plan $\mathbb{S}$, we can estimate the communication process group allocations for each parallelism strategy and sharding approach. Subsequently, this allows us to generate an executable for precise profiling using \texttt{UPro} with the estimated device meshes.
In Appendix \ref{sec:appendix_solution}, we provide a detailed workflow for solving the optimization problem presented. Note that \sysname{} identifies the \emph{top-10} strategies, and the selection of the optimal strategy is finalized through real system profiling to ensure the most effective performance.

\begin{figure*}
    \centering
    \includegraphics[width=\linewidth]{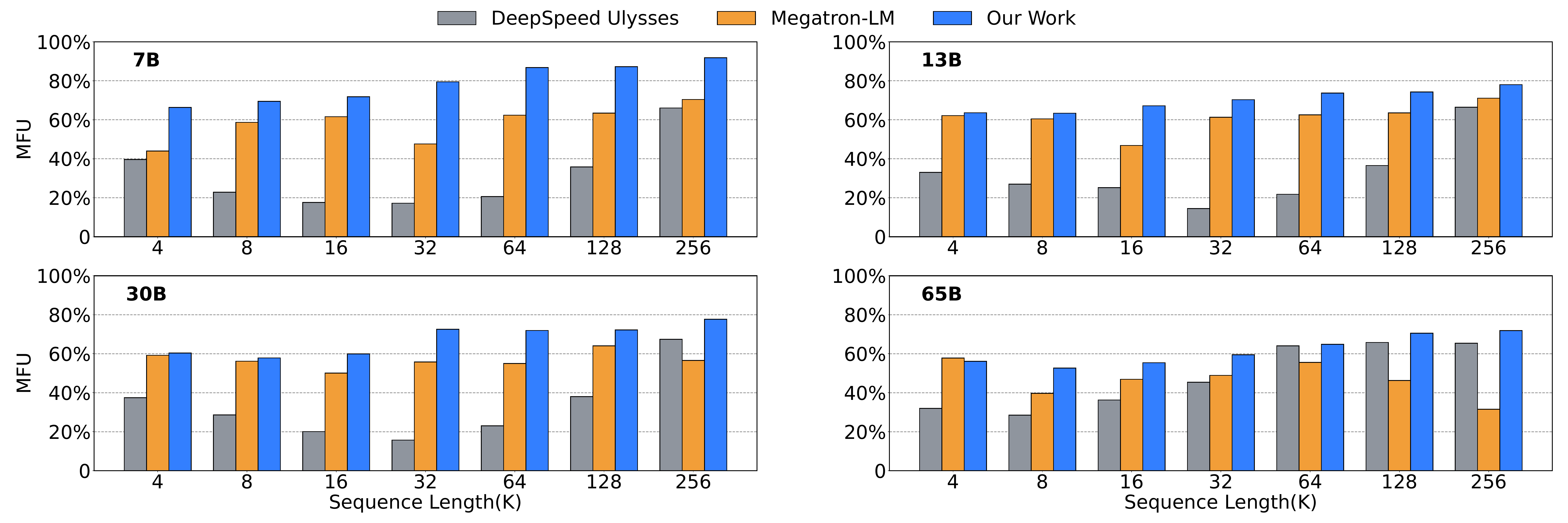}
    \caption{End-to-end evaluation results (MFU) of training different sizes of models with different sequence lengths.}
    \label{fig:ab_overlap}
\end{figure*}

%% file: sections/experiment.tex
\section{Evaluation}
\label{sec:eva}
\subsection{Experiment Setup}

\noindent\textbf{Implementation.} \sysname{} has been implemented using Python and encompasses around 35,000 lines of code (LOC). The codebase consists of 5,000 LOC for the Strategy Planner and 30,000 LOC for the Job Executor. For access to the system, please visit \emph{https://github.com/InternLM/InternEvo}. Additional insights into the implementation specifics of the Communication Scheduler and Memory Fragmentation Manager can be explored in Appendix~\ref{sec:overlap} and~\ref{sec:appendix_fragmentation}.

\noindent \textbf{Testbed.} \sysname{} undergoes a thorough evaluation in the training of Transformer-based models using the LLaMA architecture. The models encompass parameters ranging from 7 billion to 65 billion, and detailed configurations can be found in Table~\ref{tab:model_used_memory}. The training process occurs on a physical cluster comprising 16 GPU servers. Each server is equipped with 8 GPUs and 128 CPU cores, leading to a total of 128 NVIDIA Ampere GPUs. Each GPU boasts 80GB of memory. The GPUs are interconnected through NVLink and NVSwitch, while inter-node communication is facilitated by four NVIDIA Mellanox 200Gbps HDR InfiniBand.


\begin{figure*}[h]
    \centering
    \includegraphics[width=\linewidth]{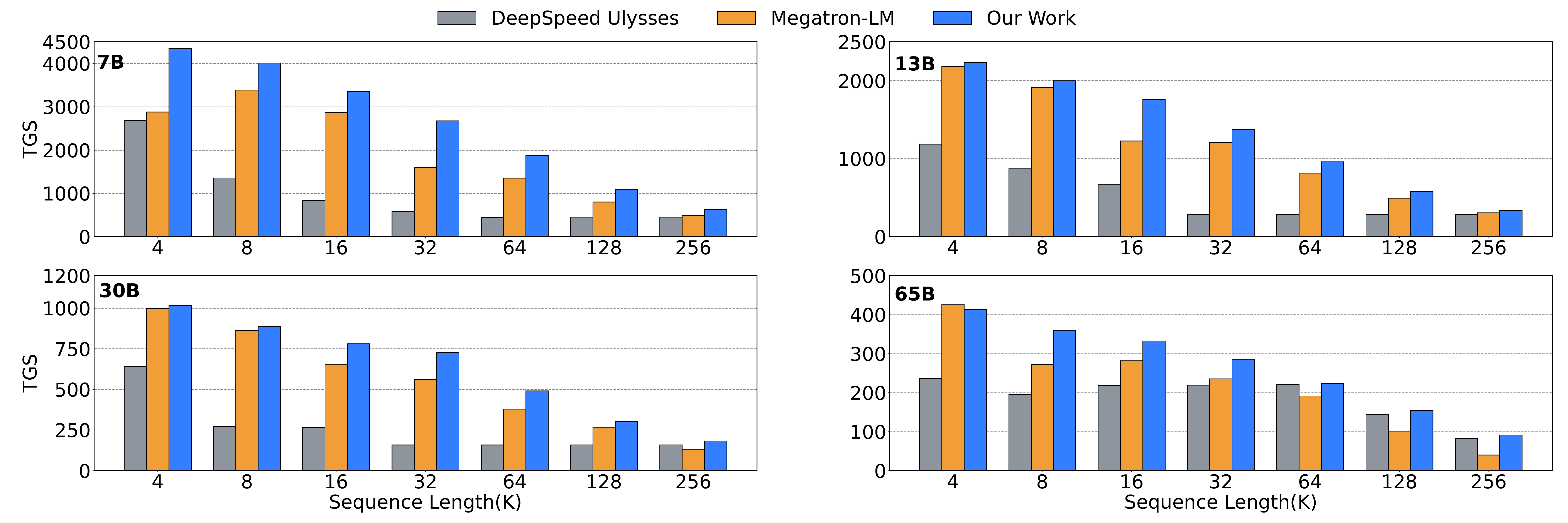}
    \caption{End-to-end evaluation results (TGS) of training different sizes of models with different sequence lengths.}
    \label{fig:end_to_end_tgs}
\end{figure*}

\noindent\textbf{Baselines and Metrics.} We benchmark \sysname{} against two state-of-the-art systems for long-sequence LLM training, namely DeepSpeed Ulysses \cite{DeepSpeedUlysses} and Megatron-LM \cite{nvidia3}. The evaluation metric is Model FLOPs Utilization (MFU) \cite{palm}. We also assess tokens per GPU per second (TGS) during training. While \sysname{} automatically searches for a parallelization strategy, we manually set the number of partitions for DeepSpeed Ulysses and Megatron-LM. The chosen strategy for these baselines is the fastest one that avoids triggering an out-of-memory error for fair comparison. It is worth noting that, in practice, the parallelization strategy selected by developers for DeepSpeed Ulysses and Megatron-LM would likely be either the same as or slower than the strategy chosen in our experiments. As a comprehensive case study, we also present the parallelization strategy identified by \sysname{}.

\subsection{End-to-End Evaluation}

\begin{figure*}[h]
    \centering
    \includegraphics[width=\linewidth]{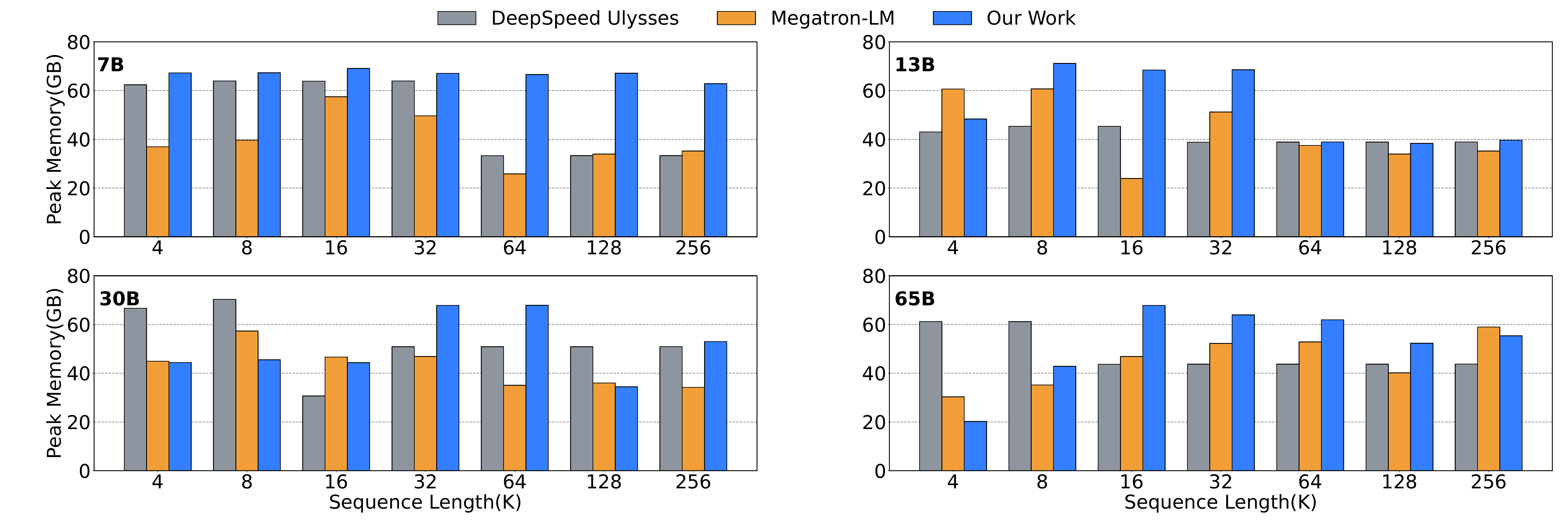}
    \caption{Peak Memory of training different sizes of models with different sequence lengths.}
    \label{fig:end_to_end_peak}
\end{figure*}

We initiate our evaluation by assessing the training speed improvement achieved by the parallelization strategy identified by \sysname{} while ensuring no out-of-memory errors. Figure~\ref{fig:ab_overlap} shows MFU when training models of varying sizes with different sequence lengths and Figure~\ref{fig:end_to_end_peak} shows the corresponding peak memory usage. For each model, we adopt a global batch size of $4M$ tokens \cite{globalbatchsize1,GShard}. In Megatron-LM and DeepSpeed Ulysses, we first tune the best micro-batch size ($b$) for each model and system configuration that maximizes the system performance. The gradients are accumulated across microbatches with micro-batch numbers ($n$). More details of the parallelism configuration we set for Megatron-LM and DeepSpeed Ulysses can be found in Appendix \ref{subsec:detailsE2E}, and the related TGS result is shown in Figure \ref{fig:end_to_end_tgs}.



When the sequence is short, the primary bottleneck in DeepSpeed Ulysses is the communication time. This is attributed to the utilization of ZeRO3 to reduce the memory requirements of model states, introducing a constant and substantial inter-node communication overhead. Additionally, the \texttt{all-to-all} communication overhead increases with the sequence length.  
Consequently, for shorter sequences ($S \leq 32K$), MFU tends to decrease as the sequence length grows.
By contrast, when the sequence is long enough, the computation time becomes the bottleneck. DeepSpeed Ulysses can benefit from the efficient \texttt{all-to-all} communication. The longer the sequence length ($S \geq 32K$), the more intensive the computation, and the higher the corresponding MFU.

In  \sysname{}, the communication overhead is minimized with hybrid parallelism and redundant sharding. Thus, MFU increases with the sequence length when training all four types of models.
Importantly, \sysname{} consistently outperforms baseline methods or achieves comparable training performance across all configurations. Notably, compared to DeepSpeed Ulysses and Megatron-LM, the achieved MFU by \sysname{} stands out, surpassing them by up to 4.8$\times$ and 2.29$\times$, respectively. This emphasizes the efficiency of \sysname{} in managing communication overhead and optimizing training performance.

When training 30B and 65B models with a sequence length of 4k, the parallelization strategy employed by \sysname{} coincides with that of Megatron-LM. As the sequence length in this scenario is relatively small, the volume of activations becomes much smaller than the model states. Consequently, the communication cost of Megatron-LM becomes sufficiently minimal, allowing it to achieve a high MFU. The parallelization strategy adopted by \sysname{} in this case aligns with the optimal strategy identified by the system, resulting in an equivalent MFU for both \sysname{} and Megatron-LM.

\subsection{Ablation Study}


\subsubsection{Analysis of Simulator}
\begin{figure}[t]
    \centering
    \includegraphics[width=\linewidth]{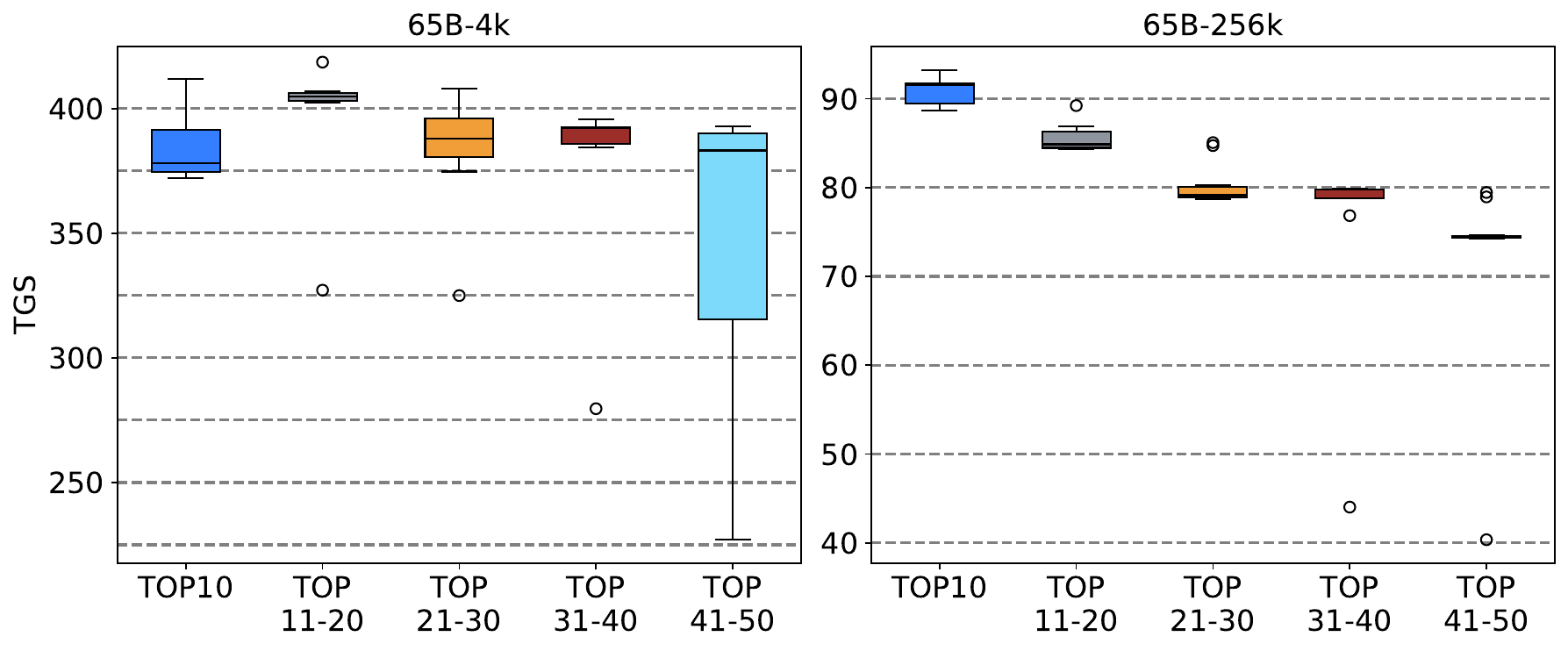}
    \caption{The performance of Top-50 configs on 65B for 4k and 256k sequence length.}
    \label{fig:abtop50}
\end{figure}

We conduct experiments to validate the effectiveness of our simulator. Treating the set of training configurations, denoted as $\mathbb{S}$, as a ranking problem, our main concern is whether the order of these configurations, especially within the top-10 solutions, is accurately selected for real-system profiling. We select the first 50 solutions generated by the simulator and group them into top 10, top 11-20, top 21-30, top 31-40, and top 41-50. The actual TGS for these 50 configurations is measured and visualized in Figure \ref{fig:abtop50}.

By comparing the simulation results for 65B-4K and 65B-256K training, we find that the 65B-256K simulation provides a more accurate estimation, indicating the simulator's good performance with longer sequences.
The observed discrepancy in the median value of top 11-20 in the 65B-4K training is due to the uncertain overlapping modeling between computation and communication, caused by frequent kernel execution and \emph{bubbles} across kernels, especially with shorter sequences. However, this discrepancy does not impact the final deployment results. The maximum value within the top 10 configurations, representing the highest TGS among the first 50 results, is captured during deployment as all top 10 configurations are executed.

\subsubsection{Analysis of Communication Scheduler}

We conduct a comparison of the TGS achieved by \sysname{} with and without the selective overlap under the same parallelization strategy. The impact of selective overlap on training the 13B model is illustrated in  Figure~\ref{fig:ab_memory}(a).

The results demonstrate that the selective overlap approach leads to improvements in TGS ranging from 0.9\% to 75.2\% across different sequence lengths. The improvement is particularly substantial when transitioning from a sequence length of 4k to 64k. In these instances, the overall training performance is notably influenced by the communication costs without overlapping.
A specific scenario exemplifying the impact of selective overlap occurs with a sequence length of 8k, where the absence of this approach results in an OOM error. This occurrence can be attributed to the system's inability to leverage the specialized memory pool dedicated to communication without selective overlap.



\subsubsection{Analysis of Memory Management}

\begin{figure}[t]
    \centering
    \includegraphics[width=\linewidth]{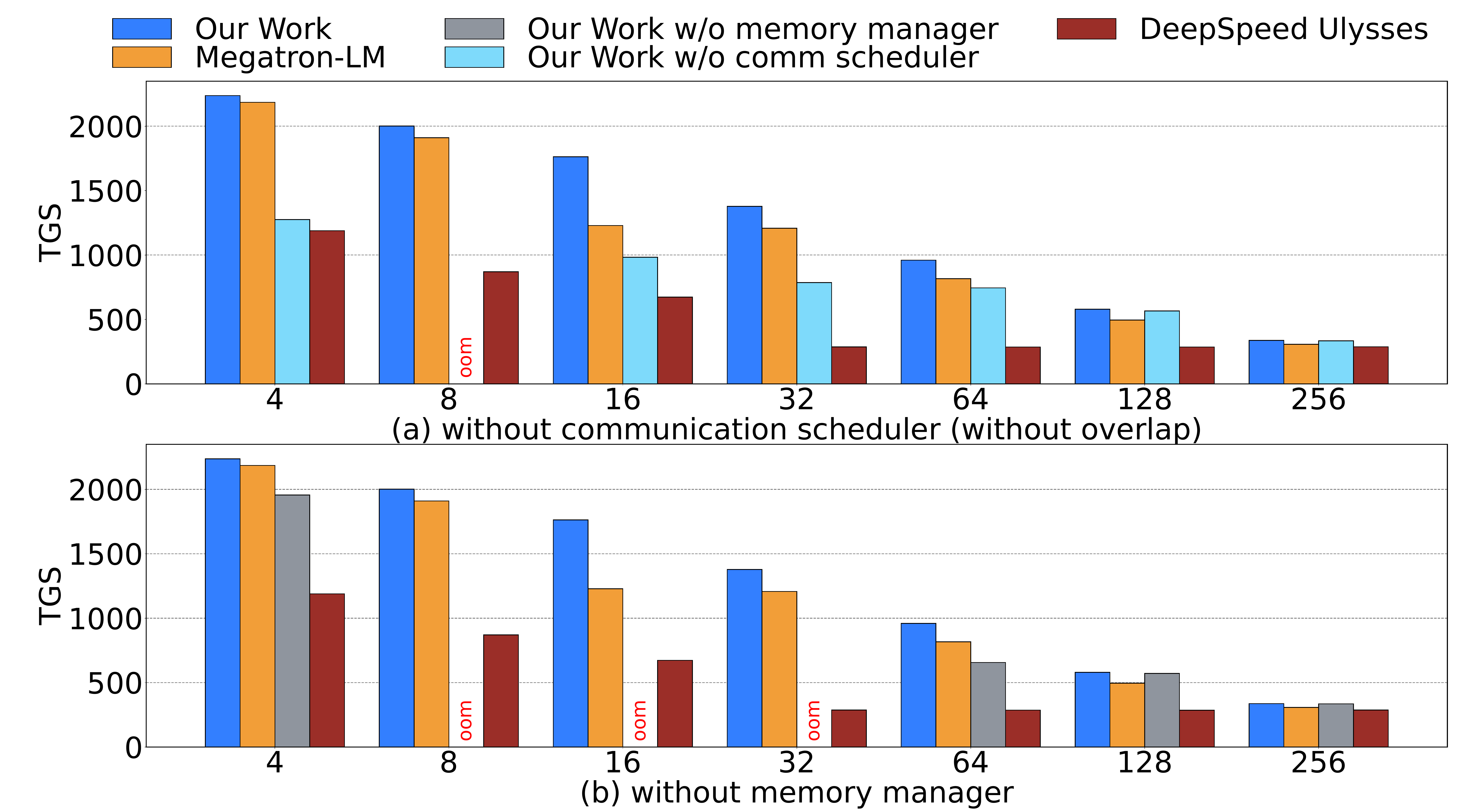}
    \caption{Effects of the Memory Management and Communication Scheduler (13B model). The $x$-axis indicates the sequence length (k).}
    \label{fig:ab_memory}
\end{figure}

In order to demonstrate the efficacy of memory management techniques, including \emph{Pinned Communication Buffer} and \emph{Memory Consolidation in Advance}, we conduct a comparative analysis of the TGS with and without the implementation of these techniques. The results are illustrated in  Figure~\ref{fig:ab_memory}(b). Without memory management, OOM errors occur when the sequence length is 8k, 16k, and 32k due to an excessive number of memory fragments. For other sequence lengths, there are no OOM errors observed, but the memory swapping (calling \texttt{CudaMalloc} and \texttt{CudaFree} due to high memory usage pressure and fragmentation) can limit the TGS.


\begin{figure*}[t]
    \centering
    \includegraphics[width=\linewidth]{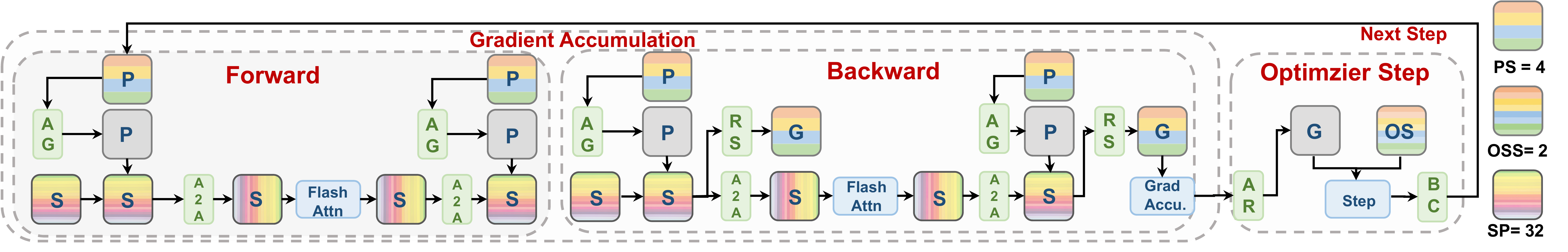}
    \caption{Visualization of parallelization strategies for training a 7B model with 256k sequence length on 128 GPUs. Different colors represent the GPUs a tensor is distributed to. Grey blocks represent a tensor that is replicated across GPUs. AG, AR, A2A, RS, and BC represent \texttt{all-gather}, \texttt{all-reduce}, \texttt{all-to-all}, \texttt{reduce-scatter}, and \texttt{broadcast} respectively.}
    \label{fig:casestudy}
\end{figure*}

\subsection{Sequence Length Scalability}

We conduct an evaluation of the largest sequence length for all three systems, configuring both $b$ and $n$ to 1 to facilitate the handling of longer sequences. To ensure compatibility with sequence parallelism ($s_{sp}$), it is crucial that the number of attention heads is divisible by $s_{sp}$. Additionally, we set the constraint $s_{dp} \cdot s_{sp} \cdot s_{pp} = N$ to guarantee the feasibility of $s_{sp}$.
For the 7B and 65B models on 128 GPUs, we set $s_{sp}$ to 32 and 64, respectively, representing their maximum $s_{sp}$ values that support longer sequences while adhering to the attention head constraint. Furthermore, we evaluate the 13B and 30B models on 80 GPUs and 96 GPUs, setting $s_{sp}$ to 40 and 48, respectively. These choices are made to ensure that the number of attention heads is divisible by $s_{sp}$.

We acknowledge that our \sysname{} is subject to limitations imposed by the number of attention heads. However, this constraint exerts comparatively less influence on our methodologies, which support sequence lengths of up to 1536K. This is in contrast to certain existing SP-3 approaches, such as those highlighted in \cite{li2023lightseq}, which primarily focus on scaling beyond the number of heads.




Table \ref{table:maxlength} provides an overview of the maximum sequence lengths supported by \sysname{} and the corresponding MFU in comparison to the baseline. \sysname{} accommodates the same sequence lengths with the baselines while maintaining a consistent 75\% MFU on average. Megatron-LM achieves an equivalent sequence length with \sysname{} for all models. This uniformity arises due to the comparable reduction in the activation memory usage by both approaches, achieved by a factor of $s_{sp}$. However, Megatron-LM incurs more network communication for activations compared to \sysname{}, resulting in slower execution.
The time-out error in DeepSpeed Ulysses occurs because, while it transmits less activation data than Megatron-LM, it also communicates model states across the world size simultaneously. Without refined communication scheduling, this frequent and resource-intensive communication triggers the time-out error in DeepSpeed Ulysses. 
\\







\begin{table}
\caption{The largest sequence length supported by \sysname{}, DeepSpeed Ulysses, and Megatron-LM (Max $S$). \XSolidBrush means time-out error.}
\label{table:maxlength}
  \begin{tabular}{@{}ccccccc@{}}
    \toprule
    & \multicolumn{2}{c}{\makecell{DeepSpeed-U}} & \multicolumn{2}{c}{\makecell{Megatron-LM}} & \multicolumn{2}{c}{\makecell{{\sysname{}}}} \\
    \cmidrule(lr){2-3} \cmidrule(lr){4-5} \cmidrule(lr){6-7}
    & \makecell{Max $S$} & MFU & \makecell{Max $S$} & MFU & \makecell{Max $S$} & MFU \\
    \midrule
    7B  & 1536K & \XSolidBrush & 1536K & 35\% & 1536K & 80\% \\
    13B & 1200K & \XSolidBrush & 1200K & 27\% & 1200K & 77\% \\ 
    30B & 960K  & \XSolidBrush & 960K  & 19\% & 960K  & 75\% \\
    65B & 512K  & \XSolidBrush & 512K  & 14\% & 512K  & 70\% \\
    \bottomrule
  \end{tabular}
\end{table}



\subsection{Case Study}

We present a visualization of the execution plan $\mathbb{S}$ obtained by \sysname{} for training a 7B model with 256k sequence length   on 128 GPUs in Figure \ref{fig:casestudy}. 
\sysname{} opts not to utilize the pipeline parallelism and activation recomputation mechanism during the training.
\sysname{} sets $s_{sp}$ as 32 to accommodate the 256k sequence on the device. It also sets $s_{oss}$ as 2 to reduce the memory usage of the optimizer state (Adam) from 42 GB to 5 GB. Given that model states of parameters and gradients constitute a small portion of overall memory usage in  7B model, \sysname{} further limits the communication costs by setting $s_{ps}$ to 4, thereby restricting the number of communication participants. \\

%% file: sections/related.tex
\section{Related Work}
\noindent\textbf{Systems for long-sequence training.} Since the memory and calculation of a single device are limited, ColossalAI-SP \cite{ColosslaiSP} first proposes the segmentation and parallelism for the sequence dimension in addition to tensor parallelism for hidden dimension and pipeline parallelism for model depth. On this basis, Ring Attention \cite{liu2023blockwise,BPT2} uses blockwise self-attention to split long sequences into different devices and overlap the communication of key-value blocks. Recently, LightSeq \cite{li2023lightseq} further improved the efficiency of long sequence modeling through load balancing for causal language modelings and a re-materialization-aware checkpointing strategy. The advantage of the above sequence parallelism is that it can break through architectural limitations and achieve infinitely long sequence modeling. In contrast, another type of sequence parallelism emphasizes maintaining the combination with existing efficient self-attention mechanisms such as FlashAttention \cite{flashatten1,flashatten2} to achieve input with almost infinity. For example, Megatron-LM \cite{nvidia3} only uses sequence parallelism during Dropout and Layernorm operations, thereby reducing activation redundancy. In addition, DeepSpeed-Ulysses \cite{DeepSpeedUlysses} uses an alternative all-to-all collective communication gathering weight for attention computation when segmenting the sequence, avoiding communication overhead that originally increases with length.

\noindent\textbf{Systems for model-parallelism training.} Parallelisms are used for distributed DNN training systems. TorchDDP \cite{torchDDP}, Horovod \cite{Horovod} support data parallelism. ByteScheduler \cite{ByteScheduler} and DeepSpeed \cite{DeepSpeed} extend data parallelism with communication and memory optimization. Tofu \cite{Tofu}, Flexflow \cite{FlexFlow}, and GSPMD \cite{GShard} leverage tensor parallelism to distribute model weight across devices. Dapple \cite{DAPPLE} and Megatron-LM \cite{Megatron-LM} also leverage pipeline parallelism to divide the model into multiple stages and distribute them across devices.

\noindent\textbf{Systems for model-sharding training.} ZeRO \cite{ZeRO,ZeRO-Infinity}, cross-replica sharding \cite{xu2020automatic} has proposed to address the memory limitation issue of traditional data parallelism strategy by sharding the model states onto all GPUs. MiCS \cite{MiCS} minimizes the communication scale on top of ZeRO for better scalability on the public cloud by partitioning model states within subgroups. FSDP's hybrid sharding \cite{PyTorchFSDP} implemented MiCS to Pytorch \cite{PyTorchDistributed}. ZeRO++ \cite{ZeRO++} takes a different approach by redundantly storing an additional set of secondary parameters on each node, in exchange for enhanced communication efficiency through parameter pre-fetching. AMSP \cite{AMSP} takes a more flexible approach by partitioning the three components of model states independently for better scalability in the training of not-so-large (7B and 13B) models on thousands of GPUs. PaRO \cite{linslike} provides a fine-grained sharding strategy for various training scenarios like parameter-efficient fine-tuning task \cite{LoRA,P-Tuningv2}.


\noindent\textbf{Automatic search for distributed training.} For general models and hardware, some systems \cite{ Whale,Piper,Alpa} employ mathematical programming techniques, such as dynamic programming and integer linear programming, to automate the configuration of parallel mechanisms. To make the immense search space more manageable, these systems simplify the problem in various ways. Flexflow \cite{FlexFlow} focuses on specific mechanisms like data and tensor parallelism, Gpipe \cite{GPipe} and PipeDream \cite{PipeDream} concentrates ZeRO on data and pipeline parallelism, and AMSP \cite{AMSP} only focus on data parallelism, Alpa\cite{Alpa} and Galvatron \cite{Galvatron} focus on all parallelism except sequence parallelism.




%% file: sections/conclusion.tex
\section{Conclusion}
In this paper, we propose \sysname{}, an automatic parallelization framework for training LLMs with long sequences. We first decouple the GPU memory management dimensions of LLM training into a brand new hierarchical space and systematically analyze their memory and communication costs. Then we develop an execution simulator to derive an efficient parallelization strategy in this hierarchical space. We design a selective overlap method and a memory fragmentation manager to further improve the computation resource utilization and memory utilization. Evaluation results show that \sysname{} can generate the optimal parallelization strategies that match or outperform existing methods in MFU.
\newpage

%% file: sections/appendix.tex
\clearpage


\appendix

\section{Appendix}

\subsection{Implementation Detail: Selective Overlap}
\label{sec:overlap}

\begin{figure}
    \centering
    \includegraphics[width=\linewidth]{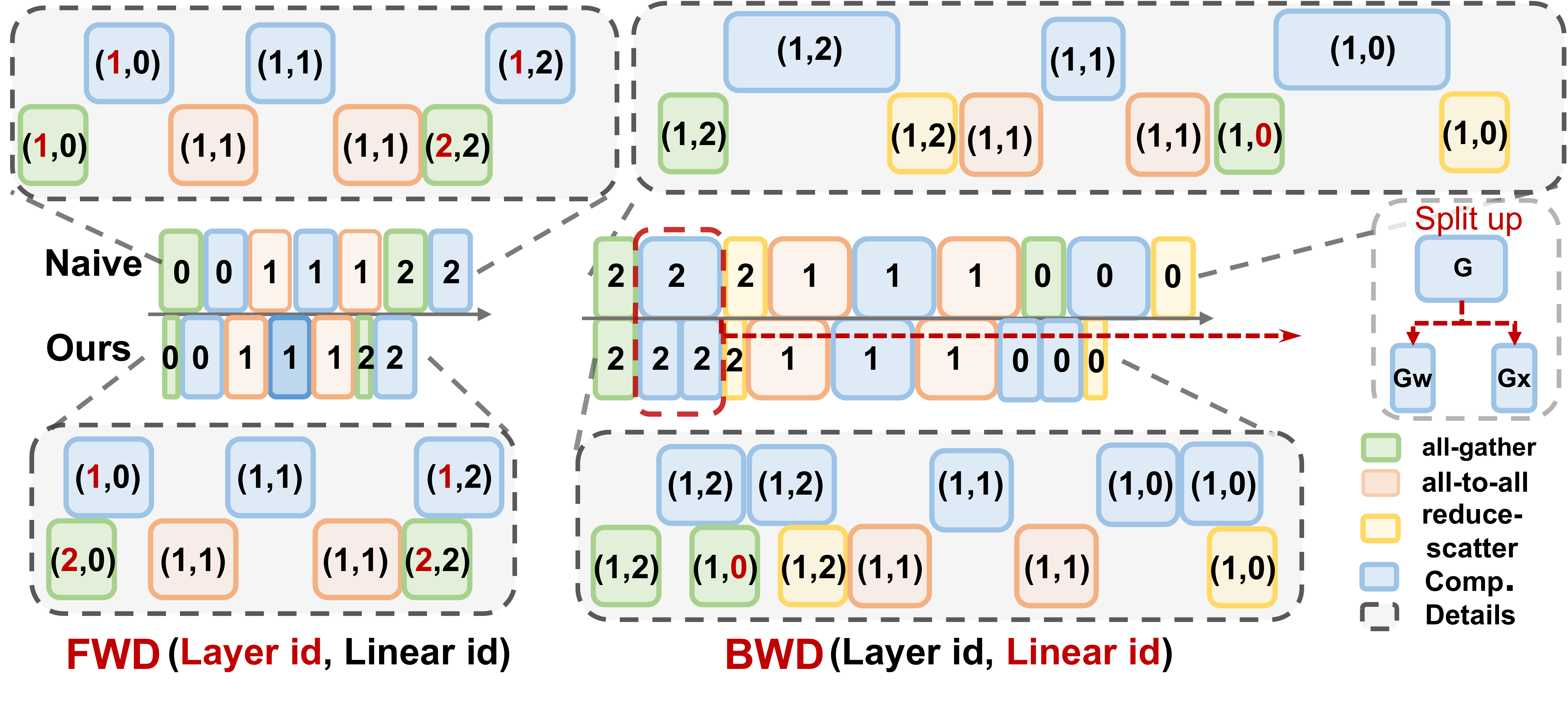}
    \caption{Comparison of \sysname's overlap with a naive pass in two layers, one attention, and three linear calculations in each layer. We use different overlap granularities in the forward and backward passes.}
    \label{fig:overlap}
\end{figure}

The Communication Scheduler module within \sysname{} introduces an innovative method designed to maximize the utilization of idle computation resources by strategically overlapping communication and computation processes. This overlap can manifest at either the inter-layer or intra-layer granularity. Although the intra-layer overlap strategy initially seems appealing due to its finer granularity and lower memory requirements for communication buffers, a closer examination reveals potential drawbacks.

A straightforward yet limited approach might involve consistently applying the intra-layer overlap strategy. However, this fine-grained strategy, while reducing memory demands, involves more frequent communication instances. In scenarios with CPU-intensive workloads, the CPU thread may struggle to timely issue every communication operator. Consequently, the cumulative latency introduced by CPU scheduling and communication may surpass the execution time of computation. This limitation indicates that intra-layer overlap alone may not consistently mask communication overhead effectively. Therefore, a judicious consideration of workload characteristics becomes crucial, necessitating an adaptive approach to choosing between inter-layer and intra-layer overlap strategies for optimizing overall system performance.

Derived from this solution is the recognition that the key to achieving our goal is to decide the overlap granularity based on actual communication latency and computation execution time. Notably, the observation that backward propagation involves computation twice as intensive as forward propagation inspires the adoption of a nuanced strategy called \textbf{selective overlap}. In this approach, a memory-efficient, fine-grained overlap is employed during backward passes, capitalizing on the longer computation duration that can tolerate greater latency. Conversely, for the forward pass, the inter-layer overlap strategy is adopted, trading off increased memory usage for a significantly smaller step time. This iterative execution of inter-layer and intra-layer communication-computation overlap aims to strike a balance between computational efficiency and memory utilization, contributing to an overall enhancement in system performance.

\textbf{Intra-Layer Overlap in Backward. }
In the backward pass, we distinguish between two computation categories: \emph{G-X} (gradient with respect to the input) and \emph{G-W} (gradient with respect to the layer's weight), as depicted in the top right corner of Figure \ref{fig:overlap}. While the conventional approach combines \emph{G-X} and \emph{G-W} into a single backward function for user-friendliness, it inadvertently introduces inefficiencies. Specifically, the \texttt{reduce-scatter} communication-related to \emph{G-W} is independent of \emph{G-X}, leading to unnecessary delays in the critical path.
To address this, we propose an intra-layer overlap strategy. This allows \emph{G-X} computations to overlap with \texttt{reduce-scatter} communication for \emph{G-W}. Additionally, we introduce backward pre-fetching, launching the next \texttt{all-gather} before the current \texttt{reduce-scatter} in the layer. This asynchronous, non-blocking task runs concurrently with \emph{G-W} computation.
An important insight guiding this strategy is that the communication order in the forward pass is the reverse of the backward pass. Leveraging this, we can anticipate which parameters to \texttt{all-gather} next in backward pre-fetching. This orchestrated approach enhances communication-computation overlap, improving overall performance during backward propagation.

\textbf{Inter-layer overlap in forward. }
Utilizing an overlap policy with module (such as linear module) granularity could result in delays for the next \texttt{all-gather} kernel to start, subsequently blocking computations on critical paths that depend on it. Such blocking effects tend to accumulate within a layer.
To address this, we adopt an inter-layer overlap strategy, aiming to mitigate the impact of kernel launch latency on forward propagation. This strategy involves obtaining parameters for the next layer in advance while computing the current layer. As depicted in Figure \ref{fig:overlap} in the forward pass, we implement a forward pre-fetch for subsequent layers. Specifically, we launch the \texttt{all-gather} for the second layer $(2,0)$ and schedule this collective communication to run concurrently with the corresponding linear module computation task of the first layer. This inter-layer overlap approach optimizes the forward pass by proactively fetching parameters for upcoming layers, reducing latency and enhancing overall efficiency.

\subsection{Implementation Detail:  Defragmentation}
\label{sec:appendix_fragmentation}

\label{sec:memory_pool}
\begin{figure}[t]
    \centering
    \includegraphics[width=\linewidth]{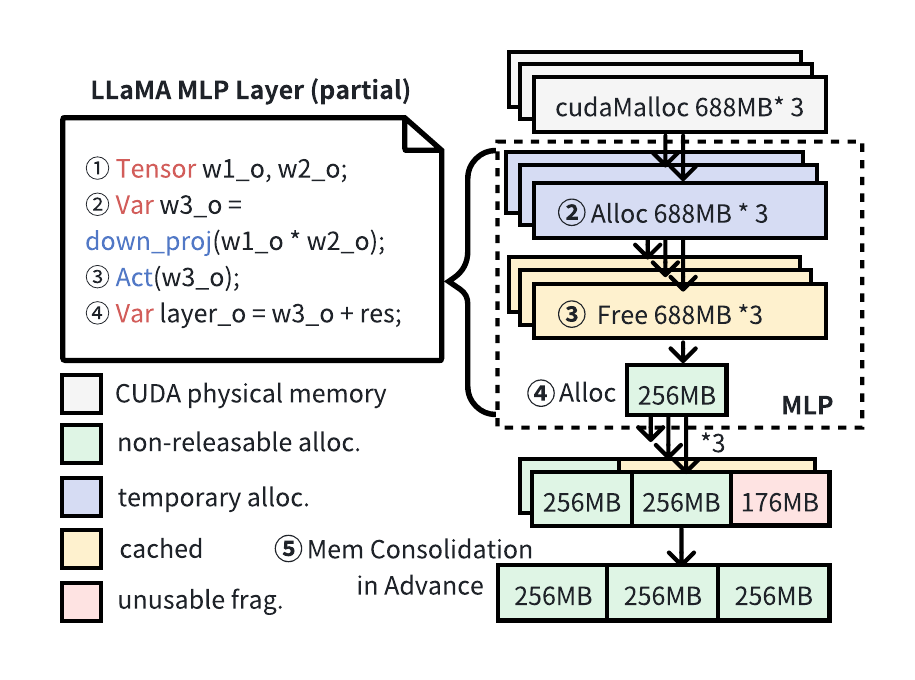}
    \caption{We illustrate the necessity of defragmentation techniques using the 65B 256K configuration from Table \ref{table:65B_config} as an example. Owing to the gate projection, up projection (\ding{172}), and the dot product of these projections (\ding{173}), there are three allocation requests of 688MiB each, with a shape of (16384, 22016) within one MLP layer. Next, the activation checkpoint mechanism releases these intermediate activations (\ding{174}); however, the MLP output is maintained as a non-releasable tensor until the corresponding backward stage (\ding{175}). As the PyTorch memory allocator employs the BestFit strategy for tensor allocation \cite{paszke2019pytorch} and the segments of 688MiB have been cached, the 256MiB MLP output is likely placed into those 688MiB segments, resulting in a fragmentation of 176MiB per segment. Such fragmentation is too small to reuse in the 65B model forward stage with a 256K sequence. In the worst-case scenario, these unrecyclable fragments could accumulate to as much as 5.32 GiB. To tackle this significant memory challenge, we proactively aggregate and flatten outputs from every three MLP layers to avert subsequent OOM (\ding{176}). Our experiments demonstrate that this anti-fragmentation approach effectively reduces fragmentation at an acceptable cost, thereby ensuring trainability.}

    \label{Fig: memory_frag}
\end{figure}

In this section, we discuss the design of \emph{Pinned Communication Buffer} and  \emph{Memory Consolidation in Advance} for reducing GPU memory fragmentation.

\textbf {Pinned Communication Buffer}. 
In long-sequence training, the continuous memory demand for activation generated by the computation stream increases. However, frequent \texttt{all-gather} communications for parameter pre-fetching may lead to GPU memory fragmentation.
To tackle this, \sysname{} introduces memory pools dedicated to \texttt{all-gather} in forward and backward passes. Before forward propagation, \sysname{} allocates a new pool with a size equal to the required size for two layers' \texttt{all-gather} communication buffer. It implements double buffer rotation, utilizing one buffer for computation and another for communication in an overlapping mode. Therefore, \sysname{} would not request memory for \texttt{all-gather} from PyTorch on the demand during training. This approach helps manage memory efficiently and addresses fragmentation concerns caused by frequent communication operations. 

\textbf{Memory Consolidation in Advance.} To prevent potential out-of-memory (OOM) issues caused by fragmentation, we adopt a strategy of consolidating small, scattered memory chunks through heuristic methods. Transformer models, especially when using the standard LLaMa model with \texttt{MLP\_Ratio 8/3} and activation recomputation, tend to experience increased internal fragmentation in MLP layers due to irregular dimensions. Our solution involves combining these irregular and small output segments to reduce fragmentation, thus preventing OOM incidents. Figure \ref{Fig: memory_frag} gives an example of such a case.
Additionally, we've recognized challenges posed by fragmented memory space, particularly occupied by gradients, especially in long sequence training. To tackle this, prior to each forward operation at every step, we proactively map the \texttt{.grad} attribute of each learnable parameter to a contiguous memory chunk.

\subsection{Solution Algorithm}
\label{sec:appendix_solution}


Algorithm \ref{algo:simulator} encapsulates the workflow employed by \sysname{} to derive optimal execution plan candidates. The execution plan $\mathbb{S}= [b, n, a, s_{pp},s_{dp},s_{tp},s_{sp},s_{ps},s_{gs},s_{oss}]$ is systematically explored, with \sysname{} considering all possible combinations of micro-batch size, micro-batch number, parallelism strategies, and sharding strategies. For a given candidate solution $\mathbb{S}_t$, \sysname{} initiates the process by estimating its memory usage and rigorously checking its adherence to constraints such as single GPU memory capacity, global batch size, and global GPU number. If $\mathbb{S}_t$ proves feasible, \sysname{} proceeds to estimate its execution time utilizing existing profiling data and default device mesh placement sequences. The algorithm dynamically maintains a curated list of the top 10 solutions, continuously updating it throughout the search iterations based on the lowest observed execution times. The final output comprises the top 10 optimal execution plans.

\begin{algorithm}
\small
\caption{Search Optimal Execution Plan}
\label{algo:simulator}
\begin{algorithmic}[1]
\State \textbf{Input:} Model Architecture, GPU Count $N$
\State \textbf{Output:} Top10 Execution Plans
\State $\mathbb{S}= [b, n, a, s_{pp},s_{dp},s_{tp},s_{sp},s_{ps},s_{gs},s_{oss}]$
\State Initialization: $\text{top10\_solutions} = \text{None}$
\State Initialization: $\text{top-10\_cost} = \infty$
\For{$s_{pp} \in \{1, 2, \ldots, N\}$}
    \For{$s_{sp},s_{tp} \in \{1, 2, 4, \ldots, {N}/{s_{pp}}\}$ } 
        \For{$s_{dp}$ \textbf{in} $\{1, 2, \ldots, {N}/{s_{pp}\cdot s_{tp}}\}$}
            \For{$b$ \textbf{in} $\{1, 2, \ldots, {B}/({S \cdot s_{dp}})\}$}
                \For{$s_{ps}$,$s_{gs}$,$s_{oss}$ \textbf{in} $\{1, 2, \ldots, N\}$}
                    \State Get micro-batch num $n={B}/({S \cdot s_{dp} \cdot b})$
                    \State Create candidate solution $\mathbb{S}_t$
                    \If{\Call{CheckConstraints}{$\mathbb{S}_{t}$}}
                        \State Estimate $T_{fwd\_bwd}(\mathbb{S}_t) + T_{update}(\mathbb{S}_t)$
                        \If{$T(\mathbb{S}_t) < \max(\text{top10\_time}$)}
                            \State Update \text{top10\_solutions} 
                        \EndIf
                    \EndIf
                \EndFor
            \EndFor
        \EndFor
    \EndFor
\EndFor
\State \textbf{Print} Best Solution: $\text{best\_solutions}$
\end{algorithmic}
\end{algorithm}

\subsection{Details of E2E Evaluation}
\label{subsec:detailsE2E}

Table \ref{tb:4} to \ref{tb:7} shows the detailed parallel configs in Section \ref{sec:eva}. Figure \ref{fig:end_to_end_peak} shows the peak memory of different sizes of models with different sequence lengths.

\renewcommand{\arraystretch}{0.9}
\begin{table*}[h]
    \small
    \centering
    \begin{minipage}{0.45\textwidth}
    \centering
    \caption{Configuration on Model 7B}
    \begin{tabular}{@{}l@{\hspace{0.12cm}}l@{\hspace{0.15cm}}c@{\hspace{0.22cm}}c@{\hspace{0.22cm}}c@{\hspace{0.22cm}}c@{\hspace{0.22cm}}c@{\hspace{0.22cm}}c@{\hspace{0.22cm}}c@{\hspace{0.22cm}}c@{\hspace{0.22cm}}c@{\hspace{0.22cm}}c@{}}
        \toprule
        \textbf{Seq} & \textbf{System} & \textbf{$n$} & \textbf{$b$} & \textbf{$s_{pp}$}  & \textbf{$s_{dp}$} & \textbf{$s_{tp}$} & \textbf{$s_{sp}$} & \textbf{$s_{ps}$} & \textbf{$s_{oss}$} & \textbf{$a$} \\
        \midrule
        \multirow{3}{*}{4k} & Megatron-LM & 8 & 2  & 1 & 64 & 2 &2 & 1 & 64 & 0 \\
        & Deepspeed-U & 4 & 2 & 1 & 128 & 1 &1 & 128 & 1 & 0 \\
        & Our Work & 4 & 2 & 1 & 128 & 1 &1 & 4 & 2 & 0 \\
        \midrule
        \multirow{3}{*}{8k} & Megatron-LM & 8 & 1 & 1 & 64 & 2 & 2 & 1 & 8 & 0 \\
        & Deepspeed-U & 16 & 2 & 1 & 64 & 1 & 2 &128 & 1 & 0 \\
        & Our Work & 4 & 1 & 1 & 128 &  1  & 1 & 4 & 2 & 0 \\
        \midrule
        \multirow{3}{*}{16k} & Megatron-LM & 4 & 1 & 1 & 64 & 2 & 2 & 1 & 64 & 0 \\
        & Deepspeed-U & 4 & 4 & 1 & 16 & 1 & 8 & 128 & 1 & 0 \\
        & Our Work & 4 & 1 & 1 & 64 & 1 &2 & 2 & 4 & 0 \\
        \midrule
        \multirow{3}{*}{32k} & Megatron-LM & 4 & 1 & 1 & 32 & 4 & 4 & 1 & 32 & 0 \\
        & Deepspeed-U & 4 & 2 & 1 & 16 & 1 &8 & 128 & 1 & 0 \\
        & Our Work & 4 & 1 & 1 & 32 & 1 &4 & 2 & 4 & 0 \\
        \midrule
        \multirow{3}{*}{64k} & Megatron & 1 & 2 & 1 & 32 & 4 &4 & 1 & 32 & 1 \\
        & Deepspeed-U & 1 & 4 & 1 & 16 & 1 &8 & 128 & 1 & 1 \\
        & Our Work & 4 & 1 & 1 & 16 & 1 &8 & 2 & 4 & 0 \\
        \midrule
        \multirow{3}{*}{128k} & Megatron-LM & 1 & 2 & 1 & 16 & 8 &8 & 1 & 1 & 1 \\
        & Deepspeed-U & 2 & 2 & 1 & 16 & 1 &8 & 128 & 1 & 1 \\
        & Our Work & 4 & 1 & 1 & 8 & 1 &16 & 2 & 4 & 0 \\
        \midrule
        \multirow{3}{*}{256k} & Megatron-LM & 4 & 1 & 1 & 16 & 8 &8 & 1 & 4 & 1 \\
        & Deepspeed-U & 1 & 1 & 1 & 16 & 1 &8 & 128 & 1 & 1 \\
        & Our Work & 4 & 1 & 1 & 4 & 1 &32 & 4 & 2 & 0 \\
        \bottomrule
        \label{tb:4}
    \end{tabular}
    \end{minipage}
    \begin{minipage}{0.45\textwidth}
    \centering
    \caption{Configuration on Model 13B}
    \begin{tabular}{@{}l@{\hspace{0.12cm}}l@{\hspace{0.15cm}}c@{\hspace{0.22cm}}c@{\hspace{0.22cm}}c@{\hspace{0.22cm}}c@{\hspace{0.22cm}}c@{\hspace{0.22cm}}c@{\hspace{0.22cm}}c@{\hspace{0.22cm}}c@{\hspace{0.22cm}}c@{\hspace{0.22cm}}c@{}}
        \toprule
        \textbf{Seq} & \textbf{System} & \textbf{$n$} & \textbf{$b$} & \textbf{$s_{pp}$}  & \textbf{$s_{dp}$} & \textbf{$s_{tp}$} & \textbf{$s_{sp}$} & \textbf{$s_{ps}$} & \textbf{$s_{oss}$} & \textbf{$a$} \\
        \midrule
        \multirow{3}{*}{4k} & Megatron-LM & 8 & 2 &  1 & 64 & 2 & 2 & 1 & 64 & 0 \\
        & Deepspeed-U & 8 & 1 & 1 & 128 & 1 &1 & 128 & 1 & 0 \\
        & Our Work & 16 & 1 & 2 & 64 & 1 &1 & 1 & 16 & 0 \\
        \midrule
        \multirow{3}{*}{8k} & Megatron-LM & 8 & 1 & 1 & 64 & 2 &2 & 1 & 64 & 0 \\
        & Deepspeed-U & 8 & 2 & 1 & 32 & 1 &4 & 128 & 1 & 0 \\
        & Our Work & 4 & 8 & 1 & 16 & 1 &8 & 128 & 1 & 0 \\
        \midrule
        \multirow{3}{*}{16k} & Megatron-LM & 16 & 1 & 1 & 16 & 8 &8 & 1 & 16 & 0 \\
        & Deepspeed-U & 8 & 2 & 1 & 16 & 1 &8 & 128 & 1 & 0 \\
        & Our Work & 8 & 1 & 1 & 32 & 1 &4 & 4  & 8 & 0 \\
        \midrule
        \multirow{3}{*}{32k} & Megatron-LM & 1 & 2 & 1 & 64 & 2 &2 & 1 & 64 & 1 \\
        & Deepspeed-U & 1 & 8 & 1 & 16 & 1 &8 & 128 & 1 & 1 \\
        & Our Work & 8 & 1 & 1 & 16 & 1 &8 & 2 & 8 & 0 \\
        \midrule
        \multirow{3}{*}{64k} & Megatron-LM & 1 & 2 & 1 & 32 & 4 &4 & 1 & 32 & 1 \\
        & Deepspeed-U & 1 & 4 & 1 & 32 & 1 &4 & 128 & 1 & 1 \\
        & Our Work & 8 & 1 & 2 & 8 & 1 &8 & 64 & 1 & 0 \\
        \midrule
        \multirow{3}{*}{128k} & Megatron-LM & 1 & 2 & 1 & 16 & 8 &8 & 1 & 16 & 1 \\
        & Deepspeed-U & 1 & 2 & 1 & 16 & 1 &8 & 128 & 1 & 1 \\
        & Our Work & 1 & 1 & 1 & 32 & 1 &4 & 16 & 1 & 1 \\
        \midrule
        \multirow{3}{*}{256k} & Megatron-LM & 1 & 1 & 1 & 16 & 8 &8 & 1 & 16 & 1 \\
        & Deepspeed-U & 1 & 1 & 1 & 16 & 1 &8 & 128 & 1 & 1 \\
        & Our Work & 1 & 1 & 1 & 16 & 1 &8 & 16 & 1 & 1 \\
        \bottomrule
        \label{tb:5}
    \end{tabular}
    \end{minipage}
\end{table*}
\renewcommand{\arraystretch}{1}

\renewcommand{\arraystretch}{0.9}
\begin{table*}[h]
    \small
    \centering
    \begin{minipage}{0.45\textwidth}
    \centering
    \caption{Configuration on Model 30B}
    \begin{tabular}{@{}l@{\hspace{0.12cm}}l@{\hspace{0.15cm}}c@{\hspace{0.22cm}}c@{\hspace{0.22cm}}c@{\hspace{0.22cm}}c@{\hspace{0.22cm}}c@{\hspace{0.22cm}}c@{\hspace{0.22cm}}c@{\hspace{0.22cm}}c@{\hspace{0.22cm}}c@{\hspace{0.22cm}}c@{}}
        \toprule
        \textbf{Seq} & \textbf{System} & \textbf{$n$} & \textbf{$b$} & \textbf{$s_{pp}$}  & \textbf{$s_{dp}$} & \textbf{$s_{tp}$} & \textbf{$s_{sp}$} & \textbf{$s_{ps}$} & \textbf{$s_{oss}$} & \textbf{$a$} \\
        \midrule
        \multirow{3}{*}{4k} & Megatron-LM & 32 & 1 & 2 & 32 & 2 &2 & 1 & 32 & 0 \\
        & Deepspeed-U & 8 & 1 & 1 & 128 & 1 &1 & 128 & 1 & 0 \\
        & Our Work & 32 & 1 & 2 & 32 & 2 &2 & 1 & 32 & 0 \\
        \midrule
        \multirow{3}{*}{8k} & Megatron-LM & 16 & 1 & 1 & 32 & 4 &4 & 1 & 32 & 0 \\
        & Deepspeed-U & 8 & 4 & 1 & 16 & 1 &8 & 128 & 1 & 0 \\
        & Our Work & 64 & 1 & 4 & 8 & 1 &4 & 1 & 8 & 0 \\
        \midrule
        \multirow{3}{*}{16k} & Megatron-LM & 2 & 4 & 1 & 32 & 4 &4 & 1 & 32 & 1 \\
        & Deepspeed-U & 2 & 8 & 1 & 16 & 1 &8 & 128 & 1 & 1 \\
        & Our Work & 64 & 1 & 4 & 4 & 1 &8 & 1 & 8 & 0 \\
        \midrule
        \multirow{3}{*}{32k} & Megatron-LM & 2 & 2 & 1 & 32 & 4 &4 & 1 & 32 & 1 \\
        & Deepspeed-U & 1 & 8 & 1 & 16 & 1 &8 & 128 & 1 & 1 \\
        & Our Work & 8 & 1 & 1 & 16 & 1 &8 & 32 & 4 & 0 \\
        \midrule
        \multirow{3}{*}{64k} & Megatron-LM & 2 & 2 & 1 & 16 & 8 &8 & 1 & 16 & 1 \\
        & Deepspeed-U & 1 & 4 & 1 & 16 & 1 &8 & 128 & 1 & 1 \\
        & Our Work & 8 & 1 & 1 & 8 & 1 &16 & 32 & 4 & 0 \\
        \midrule
        \multirow{3}{*}{128k} & Megatron-LM & 2 & 1 & 1 & 16 & 8 &8 & 1 & 16 & 1 \\
        & Deepspeed-U & 1 & 2 & 1 & 16 & 1 &8 & 128 & 1 & 1 \\
        & Our Work & 2 & 1 & 1 & 16 & 1 &8 & 32 & 1 & 1 \\
        \midrule
        \multirow{3}{*}{256k} & Megatron-LM & 2 & 1 & 1 & 8 & 16 &16 & 1 & 8 & 1 \\
        & Deepspeed-U & 1 & 1 & 1 & 16 & 1 &8 & 128 & 1 & 1 \\
        & Our Work & 1 & 1 & 1 & 16 & 1 &8 & 32 & 1 & 1 \\
        \bottomrule
        \label{tb:6}
    \end{tabular}
    \end{minipage}
    \begin{minipage}{0.45\textwidth}
    \centering
    \caption{Configuration on Model 65B}
    \label{table:65B_config}
    \begin{tabular}{@{}l@{\hspace{0.12cm}}l@{\hspace{0.15cm}}c@{\hspace{0.22cm}}c@{\hspace{0.22cm}}c@{\hspace{0.22cm}}c@{\hspace{0.22cm}}c@{\hspace{0.22cm}}c@{\hspace{0.22cm}}c@{\hspace{0.22cm}}c@{\hspace{0.22cm}}c@{\hspace{0.22cm}}c@{}}
        \toprule
        \textbf{Seq} & \textbf{System} & \textbf{$n$} & \textbf{$b$} & \textbf{$s_{pp}$}  & \textbf{$s_{dp}$} & \textbf{$s_{tp}$} & \textbf{$s_{sp}$} & \textbf{$s_{ps}$} & \textbf{$s_{oss}$} & \textbf{$a$} \\
        \midrule
        \multirow{3}{*}{4k} & Megatron-LM & 128 & 1 & 8 & 8 & 2 &2 & 1 & 8 & 0 \\
        & Deepspeed-U & 16 & 1 & 1 & 64 & 1 &2 & 128 & 1 & 0 \\
        & Our Work & 256 & 1 & 16 & 4 & 2 &2 & 1 & 4 & 0 \\
        \midrule
        \multirow{3}{*}{8k} & Megatron-LM & 64 & 1 & 4 & 8 & 4 &4 & 1 & 8 & 0 \\
        & Deepspeed-U & 16 & 2 & 1 & 16 & 1 &8 & 128 & 1 & 0 \\
        & Our Work & 256 & 1 & 8 & 2 & 1 &8 & 1 & 16 & 0 \\
        \midrule
        \multirow{3}{*}{16k} & Megatron-LM & 8 & 2 & 2 & 16 & 4 &4 & 1 & 16 & 1 \\
        & Deepspeed-U & 4 & 4 & 1 & 32 & 1 &4 & 128 & 1 & 1 \\
        & Our Work & 4 & 1 & 1 & 64 & 1 &2 & 8 & 8 & 1 \\
        \midrule
        \multirow{3}{*}{32k} & Megatron-LM & 4 & 2 & 1 & 16 & 8 &8 & 1 & 16 & 1 \\
        & Deepspeed-U & 2 & 2 & 1 & 32 & 1 &4 & 128 & 1 & 1 \\
        & Our Work & 4 & 1 & 1 & 32 & 1 &4 & 8 & 8 & 1 \\
        \midrule
        \multirow{3}{*}{64k} & Megatron-LM & 4 & 1 & 1 & 16 & 8 &8 & 1 & 16 & 1 \\
        & Deepspeed-U & 2 & 1 & 1 & 32 & 1 &4 & 128 & 1 & 1 \\
        & Our Work & 4 & 1 & 1 & 16 & 1 &8 & 8 & 8 & 1 \\
        \midrule
        \multirow{3}{*}{128k} & Megatron-LM & 4 & 1 & 1 & 8 & 16 &16 & 1 & 8 & 1 \\
        & Deepspeed-U & 2 & 1 & 1 & 16 & 1 &8 & 128 & 1 & 1 \\
        & Our Work & 2 & 1 & 1 & 16 & 1 &8 & 32 & 2 & 1 \\
        \midrule
        \multirow{3}{*}{256k} & Megatron-LM & 2 & 1 & 1 & 8 & 16 &16 & 1 & 8 & 1 \\
        & Deepspeed-U & 2 & 1 & 1 & 8 & 1 &16 & 128 & 1 & 1 \\
        & Our Work & 2 & 1 & 1 & 8 & 1 &16 & 32 & 2 & 1 \\
        \bottomrule
        \label{tb:7}
    \end{tabular}
    \end{minipage}
\end{table*}
\renewcommand{\arraystretch}{1}